\documentclass[]{aastex701}

\usepackage{booktabs}
\usepackage{multirow}

\begin{document}

\title{A Census of $^{13}$CO -traced Molecular Cloud Structures with Distances in the Northern Galactic Plane}

\author[0009-0002-2379-4395]{Shiyu Zhang}
\affiliation{Purple Mountain Observatory, Chinese Academy of Sciences, \\ 
10 Yuanhua Road, 210023, Nanjing, \\
People’s Republic of China}
\affiliation{State Key Laboratory of Radio Astronomy and Technology, Purple Mountain Observatory,  \\ 
Chinese Academy of Sciences, 10 Yuanhua Road, Nanjing 210023, \\
People’s Republic of China}
\email{syzhang@pmo.ac.cn}

\author[0000-0002-0197-470X]{Yang Su$^\dagger$}
\affiliation{Purple Mountain Observatory, Chinese Academy of Sciences, \\ 
10 Yuanhua Road, 210023, Nanjing, \\
People’s Republic of China}
\affiliation{State Key Laboratory of Radio Astronomy and Technology, Purple Mountain Observatory,  \\ 
Chinese Academy of Sciences, 10 Yuanhua Road, Nanjing 210023, \\
People’s Republic of China}
\affiliation{School of Astronomy and Space Science, University of \\
Science and Technology of China, 96 Jinzhai Road, Hefei 230026, \\
People’s Republic of China}
\email{yangsu@pmo.ac.cn}

\author[0000-0003-3151-8964]{Xuepeng Chen$^\dagger$}
\affiliation{Purple Mountain Observatory, Chinese Academy of Sciences, \\ 
10 Yuanhua Road, 210023, Nanjing, \\
People’s Republic of China}
\affiliation{State Key Laboratory of Radio Astronomy and Technology, Purple Mountain Observatory,  \\ 
Chinese Academy of Sciences, 10 Yuanhua Road, Nanjing 210023, \\
People’s Republic of China}
\affiliation{School of Astronomy and Space Science, University of \\
Science and Technology of China, 96 Jinzhai Road, Hefei 230026, \\
People’s Republic of China}
\email{xpchen@pmo.ac.cn} 

\author[0000-0001-7768-7320]{Ji Yang}
\affiliation{Purple Mountain Observatory, Chinese Academy of Sciences, \\ 
10 Yuanhua Road, 210023, Nanjing, \\
People’s Republic of China}
\affiliation{State Key Laboratory of Radio Astronomy and Technology, Purple Mountain Observatory,  \\ 
Chinese Academy of Sciences, 10 Yuanhua Road, Nanjing 210023, \\
People’s Republic of China}
\email{jiyang@pmo.ac.cn}

\author[0000-0003-4586-7751]{Qing-Zeng Yan}
\affiliation{Purple Mountain Observatory, Chinese Academy of Sciences, \\ 
10 Yuanhua Road, 210023, Nanjing, \\
People’s Republic of China}
\affiliation{State Key Laboratory of Radio Astronomy and Technology, Purple Mountain Observatory,  \\ 
Chinese Academy of Sciences, 10 Yuanhua Road, Nanjing 210023, \\
People’s Republic of China}
\email{qzyan@pmo.ac.cn}

\author[0009-0008-7695-4747]{Juan Mei}
\affiliation{Purple Mountain Observatory, Chinese Academy of Sciences, \\ 
10 Yuanhua Road, 210023, Nanjing, \\
People’s Republic of China}
\affiliation{State Key Laboratory of Radio Astronomy and Technology, Purple Mountain Observatory,  \\ 
Chinese Academy of Sciences, 10 Yuanhua Road, Nanjing 210023, \\
People’s Republic of China}
\affiliation{School of Astronomy and Space Science, University of \\
Science and Technology of China, 96 Jinzhai Road, Hefei 230026, \\
People’s Republic of China}
\email{juanmei@pmo.ac.cn}

\author[0000-0001-8060-1321]{Min Fang}
\affiliation{Purple Mountain Observatory, Chinese Academy of Sciences, \\ 
10 Yuanhua Road, 210023, Nanjing, \\
People’s Republic of China}
\affiliation{State Key Laboratory of Radio Astronomy and Technology, Purple Mountain Observatory,  \\ 
Chinese Academy of Sciences, 10 Yuanhua Road, Nanjing 210023, \\
People’s Republic of China}
\affiliation{School of Astronomy and Space Science, University of \\
Science and Technology of China, 96 Jinzhai Road, Hefei 230026, \\
People’s Republic of China}
\email{mfang@pmo.ac.cn}

\author[0000-0002-3904-1622]{Yan Sun}
\affiliation{Purple Mountain Observatory, Chinese Academy of Sciences, \\ 
10 Yuanhua Road, 210023, Nanjing, \\
People’s Republic of China}
\affiliation{State Key Laboratory of Radio Astronomy and Technology, Purple Mountain Observatory,  \\ 
Chinese Academy of Sciences, 10 Yuanhua Road, Nanjing 210023, \\
People’s Republic of China}
\affiliation{School of Astronomy and Space Science, University of \\
Science and Technology of China, 96 Jinzhai Road, Hefei 230026, \\
People’s Republic of China}
\email{yansun@pmo.ac.cn}

\author[0000-0003-2549-7247]{Shaobo Zhang}
\affiliation{Purple Mountain Observatory, Chinese Academy of Sciences, \\ 
10 Yuanhua Road, 210023, Nanjing, \\
People’s Republic of China}
\affiliation{State Key Laboratory of Radio Astronomy and Technology, Purple Mountain Observatory,  \\ 
Chinese Academy of Sciences, 10 Yuanhua Road, Nanjing 210023, \\
People’s Republic of China}
\email{shbzhang@pmo.ac.cn}

\author[0000-0002-6388-649X]{Miaomiao Zhang}
\affiliation{Purple Mountain Observatory, Chinese Academy of Sciences, \\ 
10 Yuanhua Road, 210023, Nanjing, \\
People’s Republic of China}
\affiliation{State Key Laboratory of Radio Astronomy and Technology, Purple Mountain Observatory,  \\ 
Chinese Academy of Sciences, 10 Yuanhua Road, Nanjing 210023, \\
People’s Republic of China}
\email{miaomiao@pmo.ac.cn}

\correspondingauthor{Yang Su, Xuepeng Chen}
\email[show]{yangsu@pmo.ac.cn, xpchen@pmo.ac.cn}

\begin{abstract}

We present a systematic census of $^{13}$CO-traced molecular cloud (MC) structures toward the northern Galactic plane, based on the first-phase data release of the Milky Way Imaging Scroll Painting (MWISP).  Given the optically thin nature of the $^{13}$CO line ($\langle\tau\rangle\sim0.4$) and its relatively simple position--position--velocity (p-p-$v_{\rm LSR}$) structure, we combine Gaussian decomposition with a clustering algorithm to construct a catalog of over 30,000 MC structure samples. The approach effectively mitigates severe line-of-sight projection confusion in velocity-crowded regions, particularly toward the inner Galaxy.
To anchor these clouds in three-dimensional (3D) space, we develop a novel distance-determination pipeline that integrates the background-eliminated extinction-parallax  (BEEP) method, the CU-JADE multi-jump detection algorithm, and high-precision \textit{Gaia} DR3 parallaxes. 
We performed distance measurements for approximately 3,600 MCs, and successfully derived distances for 1,751 of them. This analysis provides a comprehensive, extinction-parallax-based distance catalog for both MCs and giant cloud complexes across the northern Galactic plane. The distance uncertainties, quantified by the error propagation, establish the $<1$ kpc cloud sample as a gold standard, while confirming the robustness of structures out to 2 kpc.
The distance-resolved MC catalog reveals rich substructures within 2 kpc of the Sun, including large-scale cavities and spurs that trace the hierarchical complexity of the Galactic molecular gas. Our results suggest that Galactic rotation and massive stellar feedback jointly sculpt the gaseous disk and may regulate star formation. 
Combining multi-survey data with advanced clustering and distance-determination pipelines enables robust 3D mapping of MCs across the full four-dimensional parameter space of ($l, b, v_{\rm LSR}, d$), laying the foundation for studies of MC properties, cloud kinematics, and Galactic structure. 

\end{abstract}



\section{Introduction} 
\label{intro}
Molecular clouds (MCs) serve as the cornerstone for studying the structure and evolution of the Milky Way, acting as the crucial link connecting Galactic dynamics to the molecular gas distribution \citep{2001ApJ...547..792D, 2015ARA&A..53..583H}. As the primary reservoir of molecular gas in the Galactic plane, giant MCs constitute a significant fraction of the baryonic mass in the interstellar medium (ISM) and serve as key tracers of Galactic kinematics \citep{1997ApJ...478..624C, 2014A&A...569A.125H, 2025A&A...703A.111H}. 
Recent high-quality CO surveys have significantly advanced our understanding of the structure, distribution, and physical properties of molecular gas in the Milky Way \citep[e.g.,][]{2013ApJS..209....8D, 2017PASJ...69...78U, 2019ApJS..240....9S, 2021MNRAS.500.3064S}. In particular, data from the MWISP project have enabled systematic studies of typical giant molecular complexes \citep{2020ApJ...893...91S, 2024AJ....167..220Z}, large-scale spiral arms \citep{2024ApJ...977L..35S}, and the gas circulation within 3 kpc \citep{2024ApJ...971L...6S} of the Galaxy. These large-sample analyses, grounded in robust observational data, have provided new insights into the properties of MCs and the overall structure of our Galaxy. 

Accurate distances to MCs are fundamental for converting observed column densities and velocity information into physical quantities such as mass, virial parameter, and kinematic coupling with the Galactic potential. These properties govern scaling relations and are essential for investigating the dust-gas relation in diffuse regimes and core mass functions in dense environments \citep{1987ApJ...319..730S, 2010ApJ...721..686P, 2017MNRAS.471.3494Z, 2010ApJ...724..687L}. Previous distance estimates for MC catalogs based on kinematic distances and machine-learning methods \citep[e.g., SEDIGISM, FUGIN;][]{2021MNRAS.500.3027D, 2023PASJ...75..279F} effectively alleviate the large-scale near/far distance ambiguity in the inner Galaxy, yet the physical association between gas and nearby objects in the Solar neighborhood remains uncertain. While these surveys provide distance measurements for a substantial number of MCs, our work employs the gas--dust extinction correlation to reduce distance uncertainties for targets within $\sim 3$~kpc, which constitutes a distinct approach to resolving the MC distance problem in the Solar neighborhood. By measuring precise distances to a large sample of MCs, particularly those with crowded velocities toward the inner Galaxy, we construct their three-dimensional (3D) distribution and kinematics from a bottom-up perspective, thereby revealing rich structural and kinematic features of the local molecular gas.   

Historically, MC distances have been inferred primarily via at least four methods: kinematic distances based on Galactic rotation models, trigonometric parallaxes of associated masers \citep{2006Sci...311...54X, 2019ApJ...885..131R}, photometric/parallax-inverted distances derived from association with young stellar objects \citep[e.g.,][]{1965gast.book.....B, 1999AJ....117..354D}, and extinction-based methods \citep{1923AN....219..109W}. However, kinematic distances suffer from severe degeneracy and systematic uncertainty, particularly in the inner Galaxy where the near-far ambiguity is pervasive. In addition, in the longitude range near the tangent points such as $l\sim 70^{\circ}-100^{\circ}$ and $l\sim 160^{\circ}-200^{\circ}$ in the anti-Galactic-center, the kinematic distances are degenerate with the observed velocities, and small uncertainties in the adopted rotation curve can lead to large uncertainty in the derived distances. These limitations severely compromise the reliability of kinematic distances for detailed studies of MCs in the solar neighborhood \citep{2020ApJ...898...80Y, 2022ApJ...925..201P}. 

Masers observed for trigonometric parallaxes in the sample of \citet{2019ApJ...885..131R} are predominantly associated with high-mass star-forming regions (HMSFRs). Consequently, they offer limited constraints on the gas distribution, especially for the nearby, relatively low-mass MCs that dominate the population. Moreover, masers with precise parallaxes are exceedingly rare and, as point sources, they do not adequately trace the distances to extended MCs, whose depth can vary significantly along filamentary structures \citep{2018A&A...619A.106G}. Before Gaia, MC distances which relied on associations with young stellar objects or extinction jump techniques were largely restricted to nearby star-forming regions (e.g., Taurus, Perseus, and Orion) due to the scarcity of high-precision stellar parallaxes. No method alone can effectively constrain the majority of clouds in the solar vicinity. 

The synergy between large sample \textit{Gaia} parallaxes and stellar reddening has effectively mitigated these limitations. Pioneering works by \citet{2019A&A...624A...6Y}, \citet{2019ApJ...879..125Z}, and \citet{2020MNRAS.493..351C} employed the distance-extinction jump method to measure distances to hundreds of MCs. Subsequent studies, such as \citet{2019ApJ...885...19Y}, \citet{2024A&A...685A..39M}, and \citet{2024AJ....167..220Z}, further incorporated CO velocity information to disentangle clouds along the line of sight (LOS). They also derived physical parameters such as radial velocity and linewidth. Consequently, studies of MCs within 3 kpc of the Sun have progressed toward characterizing their precise four-dimensional ($p$-$p$-$v_{\rm LSR}$-$d$) properties.

Despite these advances, existing distance-resolved MC catalogs are still constrained by limited sample sizes, incomplete coverage, and insufficient sample homogeneity. The first limitation arises from the scarcity of large-area surveys providing sub-arcminute-scale resolution. The second stems from the lack of effective measurement techniques for complex, crowded regions along the Galactic plane, particularly toward the inner Galaxy. Recent efforts by \citet{2024AJ....167..220Z, 2025AJ....170..185Z} attempted to measure distances to $^{13}$CO clouds in velocity-crowded, high-extinction regions toward Cygnus. By separating MCs in the Cygnus Rift from those in background star-forming regions, these works reduced complexity through velocity decomposition and multi-extinction-jump detection. This strategy improved distance completeness, although these studies remained confined to specific targeted regions.

With the release of the MWISP first-phase data, \citet{2026ApJS..284...71M} systematically measured distances for 1573 clouds based on a DBSCAN-identified $^{12}$CO catalog containing over 100,000 entries, achieving unprecedented sample size and accuracy. Nevertheless, distance completeness in the first Galactic quadrant remains low, significantly hampered by the impact of velocity crowding on cloud identification and distance measurements. Thus, a global and complete picture of the molecular gas distribution and structure in the solar neighborhood remains elusive.

In the crowded velocity fields of the Galactic plane, LOS blending introduces significant uncertainties, leading to varying degrees of contamination in the derived distances. Consequently, the effective disentanglement of MC members from unrelated dust components is critical to resolving this issue \citep{2019ApJ...885...19Y}. These inversion techniques, combining cloud decomposition and jump point analysis, are essential for isolating individual cloud structures and eliminating distance contamination \citep{2019MNRAS.485.2457H, 2019A&A...628A..78R, 2025AJ....170..185Z, 2025ApJ...994...76L}. Compared to $^{12}$CO, $^{13}$CO is generally optically thinner and traces column density more reliably. Although recent studies reveal significant internal substructure within $^{13}$CO clouds \citep{2024AJ....167..220Z}, the spectral profiles of these MCs remain relatively simple, allowing them to be distinguished even in velocity-crowded regions and thus minimizing LOS confusion. 

While previous efforts successfully applied similar techniques to limited regions (e.g., Cygnus), this work extends such studies to the entire northern Galactic plane. Specifically, we employ Gaussian decomposition and multi-jump point detection to accurately decompose cloud and extinction profiles, thereby mapping the 3D distribution and velocity field of molecular gas in the solar neighborhood across a wide range of spatial scales. 
This work aims to address the difficulty of kinematic distance determination along crowded sightlines in the inner Galactic plane, and provides a large and relatively homogeneous set of extinction-based distances. These distances are tied to velocity-decomposed MWISP $^{13}$CO ($J=1\text{--}0$) structures, and cover MCs in the crowded inner-Galaxy regions. 

The structure of this paper is as follows. In Section \ref{sec:sec2}, we describe the observational data from the MWISP survey and outline our methodology, including the multi-jump point detection algorithm. Section \ref{tech} addresses the challenge of LOS confusion, detailing the identification and mitigation of multi-component contamination in both velocity and distance space. We present the resulting MC catalog in Section \ref{catalog}, followed by an analysis of the 3D distribution and kinematics of the solar neighborhood in Section \ref{sect5}. Finally, we discuss the implications for Galactic structure in Section \ref{discuss} and summarize our findings in Section \ref{conclusion}.

\begin{figure*}[ht!]
\plotone{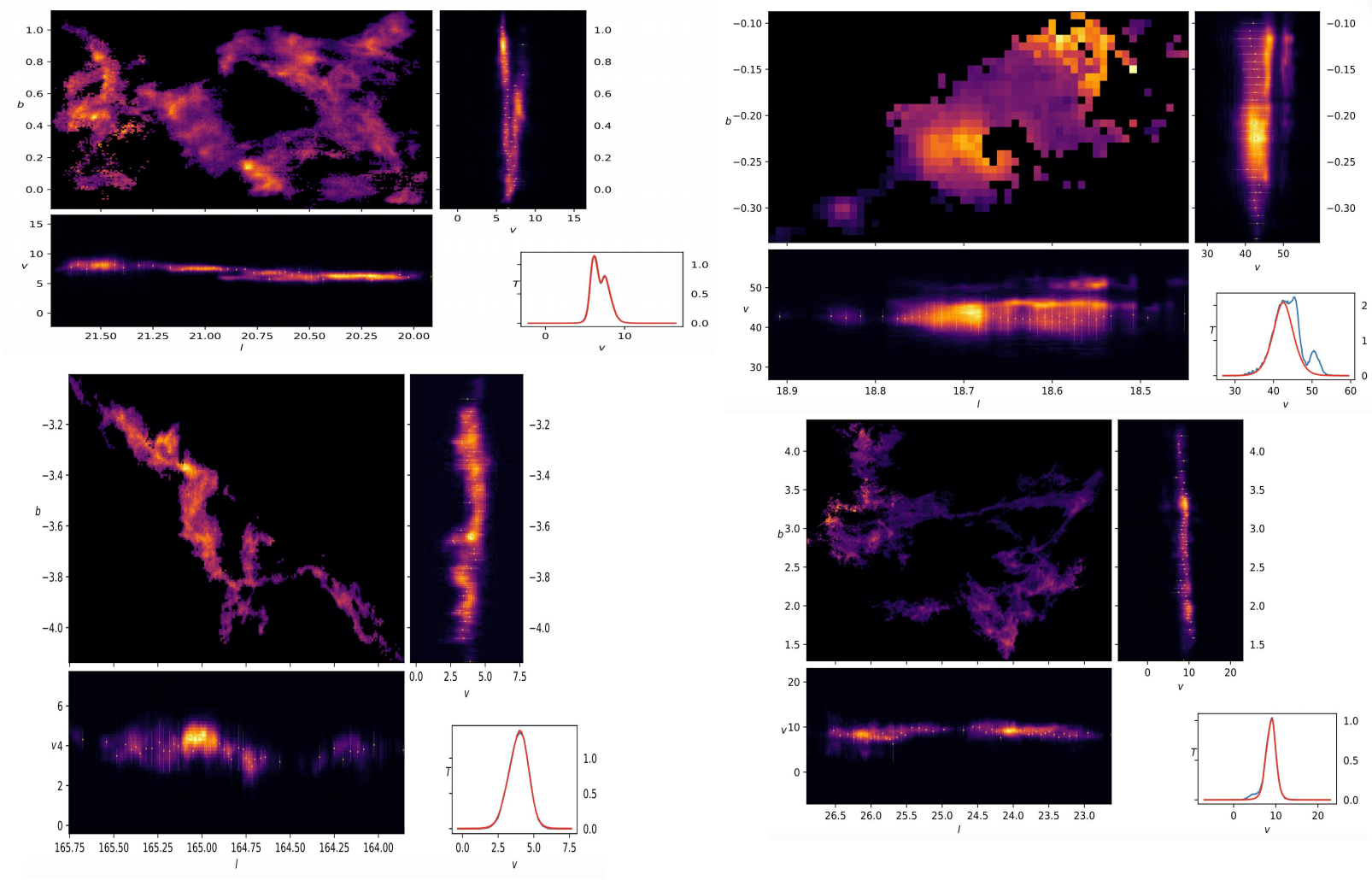}
\caption{ 
Examples of four identified $^{13}$CO MC structures.
For each sample, the colored panels display the integrated intensity map, the longitude-radial velocity $l-v$ diagram, and the latitude-radial velocity $b-v$ diagram, respectively. Note that we overlay the cloud structures extracted from the raw data onto the masks defined by the identified cloud boundaries \citep[see details in][]{2024AJ....167..220Z}. The subplots in the lower right corners show the averaged spectra within the rectangular regions. The blue curves represent the raw data, while the red curves denote the Gaussian decomposition results selected via cloud clustering to match the target cloud. The fitted centroid velocities and line widths are overplotted in the $l-v$ and $b-v$ diagrams to delineate the kinematic structure of the target cloud.
\label{fig:fig1}}
\end{figure*}

\section{Data and Technologies} \label{sec:sec2}

\subsection{Data}
\label{data}
This study utilizes data from the first phase of the Milky Way Imaging Scroll Painting survey \citep[MWISP;][]{2026ApJS..282...65Y}, covering a Galactic longitude range of $l=10^{\circ} \sim 230^{\circ}$ and a latitude range of $b=\pm5^{\circ}\!.25$. The observations achieve an angular resolution finer than 1 arcminute ($1'$) combining a 9 beams superconducting receiver and a 13.7m single dish telescope (PMO 13.7). The survey data include three CO isotopologue lines:
$^{12}$CO ($J=1\text{--}0$), $^{13}$CO ($J=1\text{--}0$), and C$^{18}$O ($J=1\text{--}0$). The $^{13}$CO ($J=1\text{--}0$) data are sampled at a velocity resolution of $\sim 0.17{\rm ~km~s^{-1}}$, with a typical sensitivity of better than 0.3 K.
The raw data were regridded into a regular 3D ($l$--$b$--$v$) data cube with a spatial pixel size of $30''$ to facilitate subsequent analysis. We filter the data to include only the region $b=\pm5^{\circ}\!.15$, excluding the low-S/N boundaries of the observational units. The final mosaicked data yield a comprehensive and unbiased view of the molecular gas distribution in the northern Galactic plane. 

The precise parallaxes and stellar reddening are all from Gaia DR3 sources \citep{2023A&A...674A...1G}. In other words, we choose reliable pure Gaia photometry and astrometric solutions ($A_{G}$-parallax) within the same sky coverage above. We adopt the extinction parameter $A_G$ from the GSP-Phot catalog. The reddened spectral energy distributions are integrated over the Gaia $G$ passband to derive magnitudes, which can then be compared with the corresponding unreddened values \citep{2023A&A...674A..27A}. Geometric distances are estimated from parallaxes using a simple Monte Carlo inversion for stars within 5 kpc. The uncertainties in $A_G$ and parallax are approximately 0.06 mag and 0.1 mas, respectively. 
The data columns and selection criteria are the same as in \citet{2024AJ....167..220Z}.

\subsection{Gaussian Decomposition and Clustering Techniques} \label{method b} 

Identifying MC structures in the velocity-crowded inner Galaxy is essential for connecting the physical properties of molecular gas to star formation and Galactic kinematics. \citet{1986ApJ...301..398M} and \citet{1989ApJ...339..149S} showed that most first-quadrant molecular gas resides in identifiable giant MCs, confirming them as the primary sites of star formation. Robust cloud decomposition in complex regions is therefore required to accurately trace gas kinematics, link cloud-scale structure to star formation, and constrain the large-scale structure of the Milky Way.

Complexities in molecular line data, particularly ambiguous cloud boundaries and velocity confusion, pose significant challenges for cloud identification. Cloud definitions vary widely depending on resolution, sensitivity, algorithm principles and parameter settings, reflecting the complex, fractal, and hierarchical nature of molecular gas \citep{2016ApJ...822...52R, 2021ApJS..257...51Y, 2025MNRAS.539.3617Y}. Recent progress in velocity decomposition algorithms based on Gaussian decomposition, such as ACORNS \citep{2019MNRAS.485.2457H}, ISMGCC \citep{2024RAA....24k5005F}, and GDCluster \citep{2025ApJ...994...76L}, has enabled the dissection of crowded regions like the CMZ and the Galactic tangent point. These methods outperform traditional PPV clustering, including Dendrogram \citep{2008ApJ...679.1338R}, SCIMES \citep{2015MNRAS.454.2067C}, and DBSCAN \citep{2020ApJ...898...80Y}, in velocity-crowded regions. The cloud decomposition method achieve this by preventing the erroneous aggregation of distinct clouds with similar radial velocities, a common issue in the inner Galaxy that leads to artificial cloud structures. 

We adopt $GaussPy+$ \citep{2019A&A...628A..78R} as our primary Gaussian decomposition algorithm. We selected two representative subregions to train the smoothing parameters:
one toward the inner Galactic plane at $l \approx 30^\circ$, and the other toward
the outer Galactic anti-center direction, spanning Galactic quadrants II and III.
The initial smoothing parameters were set to $(2.0,\,6.0)$, and the GaussPy+
training module was applied to the two regions separately.
After approximately 60 and 30 iterations, respectively, the $F1$-score \citep[e.g.,][an evaluation metric used to measure the performance of a classification model]{2025AJ....170..185Z} converged
in both cases. The training result reveals that the inner Galactic region requires larger smoothing parameters, whereas areas in the second and third quadrants are better suited to smaller values.
Consequently, we apply two sets of smoothing parameters. For Galactic longitudes $10^{\circ}<l<130^{\circ}$
, the parameters for narrow and broad components are ($\alpha1, \alpha2$)=(2.24, 4.17), respectively. For $l>130^{\circ}$, we adopt ($\alpha1, \alpha2$) =(1.75, 3.35). The choice of optimal smoothing parameters may depend on the local environments (e.g., density, velocity dispersion, and LOS crowding). A detailed discussion of their optimization will be presented in our subsequent work on $^{12}$CO cloud. 

We set the peak intensity threshold for fitting at $5\sigma$, where $\sigma$ corresponds to the noise level of the smoothed spectrum. Compared with the noise in the original data, the rms level after smoothing is lower, about 0.24 K, resulting in a peak intensity cutoff of about 1.2 K. This criterion leaves a considerable number of faint $^{13}$CO emission features un-fitted. Manual inspection indicates that the cloud structures identified under these parameters are reasonably reliable (see examples in Figure \ref{fig:fig1}).
Ultimately, we obtain 4,179,016 Gaussian components from 3,156,264 pixels, yielding a spatial filling factor of approximately 9.67\% for the whole coverage of MWISP phase I data. The maximum number of Gaussian components per pixel is 12. The fitted line widths (FWHM) are approximately $\rm 2.00^{+2.36}_{-0.92}~km~s^{-1}$, with peak amplitudes reaching up to 28.77 K. 

For the $^{13}$CO velocity components obtained via Gaussian decomposition, we employ the ACORNS algorithm as the clustering method to identify cloud structures and generate the cloud catalog. We execute the program using the default parameter set and adopt the modified version from \citet{2024AJ....167..220Z} to handle missing velocity components. For the survey-wide data, we made no additional modifications to the algorithm. It is worth noting that scaling the data volume from sub-regions to the entire northern Galactic plane leads to a nonlinear increase in computational load, resulting in exceptionally slow processing. Fortunately, for the relatively compact and sparse $^{13}$CO structures, the entire workflow completes within approximately six months in our computing resource allocation (i.e., nodes with dual 16-core 2.8 GHz AMD EPYC 7302 CPUs and 4 TB of memory.).

Ultimately, we identify 30,626 cloud structures containing nine or more pixels (where one beam corresponds to approximately four pixels). 
In Figure \ref{fig:fig1}, we present the intensity, velocity, and morphological information of four 
$^{13}$CO cloud structures. Highly structured features with complex velocity profiles are evident in both the inner and outer Galaxy. The average spectra demonstrate that our workflow effectively decomposes and identifies cloud structures, performing robustly even in velocity-crowded regions (top-right panel of Figure \ref{fig:fig1}).
A total of 4,042,581 Gaussian components were clustered into the final cloud catalog, accounting for 96.74\% of all fitted components. The statistical distribution of optical depth for all $^{13}$CO components yields $\tau = 0.38^{+0.20}_{-0.12}$, with 97.88\% of the components having $\tau < 1$. The largest structure is located in the direction of Cyg OB7, comprising 120,030 pixels. For any given cloud structure, each pixel contains only a single Gaussian velocity component. All cloud structures fall within the velocity range of [-96.15, 171.43] $\rm km~s^{-1}$, with mean column densities ranging from 3.65 to 596.84 $\rm \times 10^{20} cm^{-2}$ under the local thermodynamic equilibrium (LTE) condition by adopting empirical $\rm H_2$/$^{13}$CO abundance ratios \citep[$\sim8\times10^{5}$;][]{1982ApJ...262..590F, 1998AJ....116..336N, 2005ApJ...634.1126M}.

\subsection{BEEP and CU-JADE Method}
\label{method a}
Regarding the distance determination, we employ the background-eliminated extinction-parallax \citep[BEEP;][]{2019ApJ...885...19Y} and CUSUM-based Jump point Analysis for Distance Estimation \citep[CU-JADE;][]{2025AJ....170..185Z}. These methodologies are extensively described in \citet{2019ApJ...885...19Y} and \citet{2024AJ....167..220Z, 2025AJ....170..185Z}. We briefly outline the underlying principles and performance characteristics below. 

\begin{figure*}[ht!]
\plotone{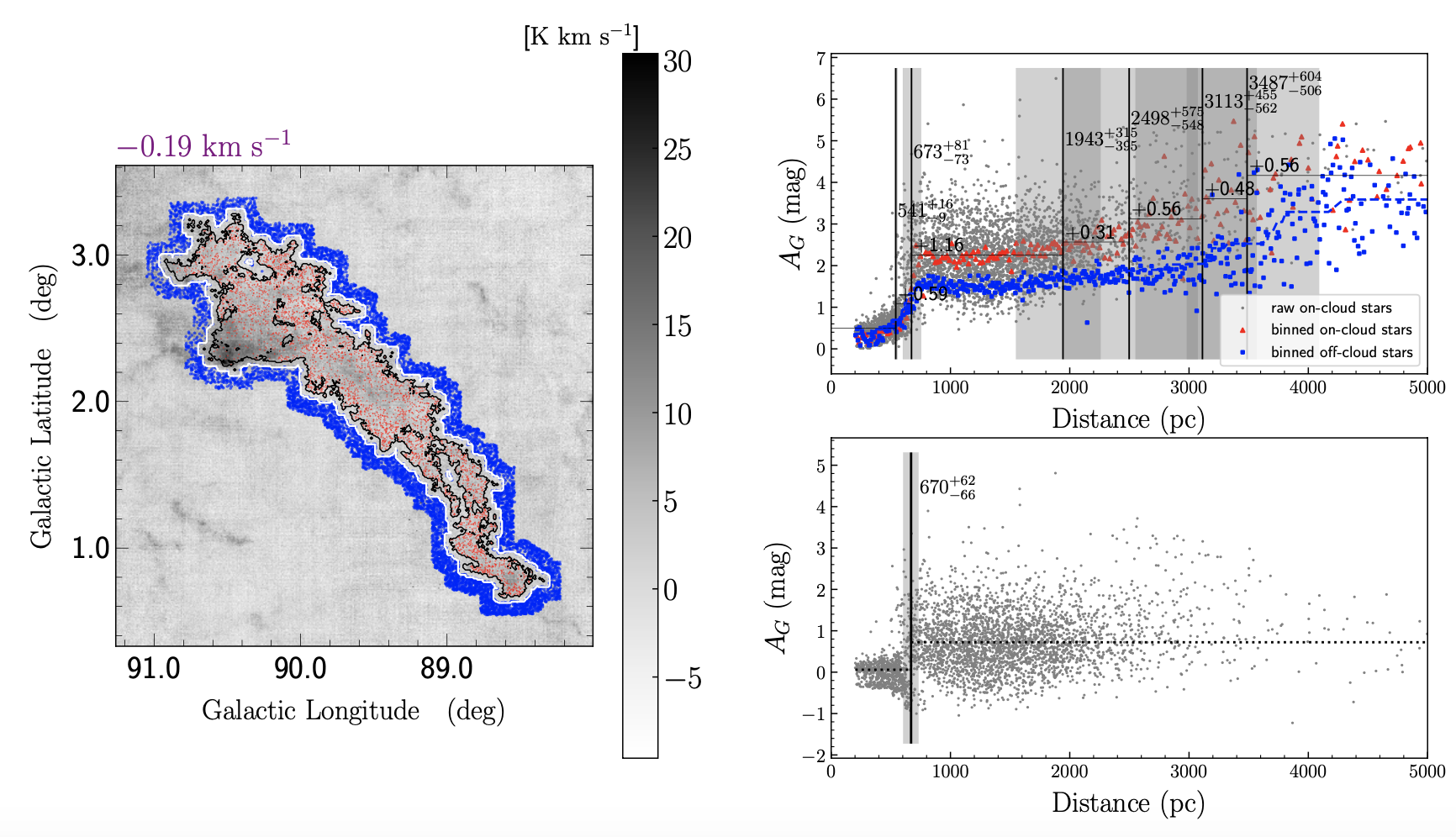}
\caption{ 
Distance measurement of a $^{13}$CO cloud structure toward Cygnus OB7 using CU-JADE and BEEP. 
\textit{Left panel:} Integrated intensity map of the target cloud. The solid black contour delineates the cloud boundary from ACORNS. Red dots denote the selected on-cloud stars within the contour, while blue dots represent field stars selected by extending outward from the boundary, spanning from the white line at $2'$ to $8'$. The centroid velocity of the cloud is indicated in the upper left corner.
\textit{Top-right panel:} Results from the CU-JADE detection. Significant extinction components are concentrated in the solar neighborhood ($<1$ kpc) and the inter-arm region ($\gtrsim 2$ kpc). Gray points represent individual on-cloud stars, red triangles show the binned averages of on-cloud stars in $20$ pc intervals, and small blue squares indicate the binned averages of field stars in $20$ pc intervals. The detected jump distances (with error bars) and corresponding extinction increments are labeled.
\textit{Bottom-right panel:} Results after applying the BEEP procedure. The jump point identified by CU-JADE corresponds to the location of the target cloud. The final adopted distance is $673$ pc, as indicated in the top right panel. 
\label{fig:fig2}}
\end{figure*}

BEEP is designed to highlight extinction jumps induced by target clouds along the LOS, making it particularly suitable for distance measurements in the low-latitude Galactic plane.
Our approach begins by defining the cloud boundaries to identify ``on-cloud" stars within the cloud contours.
These on-cloud stars trace the positions of all extinction jumps toward the target cloud. In the low-latitude regions, severe projection confusion implies that a single LOS often contains multiple extinction components distributed in front of or behind the target cloud, totaling up to ten or more distinct features \citep{2024ApJ...973..136O, 2025AJ....170..185Z}. Complementarily, ``off-cloud" stars (field stars later) are selected by extrapolating the cloud boundaries, under the assumption that both on-cloud and field stars share a similar baseline extinction-distance relationship, except for the discontinuity introduced by the target cloud.
We fit the extinction-distance curve of the field stars to establish the baseline. Subtracting this baseline from the observed extinction of the on-cloud stars isolates and highlights the signature of the target cloud (Figures \ref{fig:fig2} and \ref{fig:fig2_2}). Ideally, this subtraction yields a single, clean extinction jump. In practice, however, residual jumps from other components persist, although the jump corresponding to the target cloud remains the most prominent feature. 

The CU-JADE method differs fundamentally from traditional single-jump-point models. Its primary function is to explore the full range of plausible distance solutions. CU-JADE identifies jump points by evaluating their statistical significance relative to the dispersion of the extinction values, iterating through all potential discontinuities in the CUSUM chart via a segmentation process. Crucially, this model inherently rejects negative jump points and does not require a priori specification of the number of jumps. The significance of a jump is determined solely by the raw data. Extensive testing confirms its robustness in detecting jump points within extinction-distance space.

Single-jump-point models are of limited utility in the complex environment of the Galactic plane. They fail to capture the actual distribution of multiple cloud components along the LOS and introduce substantial fitting biases. Our analysis reveals that single-jump models typically recover the distance corresponding to the strongest extinction jump, which often does not coincide with the target cloud. Furthermore, these models suffer from accuracy degradation. For instance, in cases of closely adjacent jumps (e.g., $\sim100$ pc), a single-jump fit tends to converge to an intermediate value between the two true distances, introducing extraneous errors (see Section \ref{tech} and Appendix \ref{appa}).
To address the challenge posed by densely packed extinction components in the inner Galaxy, we present the characteristics of our sample in the following sections and detail the application of CU-JADE in conjunction with the BEEP method.

\section{Multiple Extinction jumps and Contamination Elimination}
\label{tech}

Spatially and kinematically overlapping components induce a many-to-many relationship between velocity structures and extinction features, resulting in erroneous target jump identification. And the accuracy of jump localization and error characterization directly govern the reliability of the final distance. Traditional single-jump methods often omit both scenarios. For instance, numerous cases exist where closely projected components result in strongest, centered, tending toward the mean distance assignments.

We quantify the complexity of the LOS structures by counting the number of velocity components (particularly in the inner Galaxy), alongside the number of strong extinction features. Independent velocity measurements can assign the same extinction jump to different PPV structures, producing spatially clustered cloud substructures with similar distances. Excluding contaminated components relies critically on the reliability of structural decomposition and multi-jump analysis. This affects the robustness of field star selection and background subtraction, as well as the completeness and accuracy of multi-jump point detection. 

We use the CU‑JADE algorithm to detect extinction jumps in the extinction–distance relation. For the unprocessed stellar sample, CU‑JADE identifies monotonically increasing extinction features along the LOS (e.g., upper right panel in Figure \ref{fig:fig2}). For the BEEP‑processed sample, ideally only one significant jump remains, corresponding to the target cloud. Given the uncertainties in the BEEP fitting, we match this jump to the true target jump from the unprocessed sample within their $1\sigma$ uncertainty. If multiple comparable jumps remain after BEEP, we adopt the one with the largest extinction increment as the target. 

\begin{figure*}[ht!]
\plotone{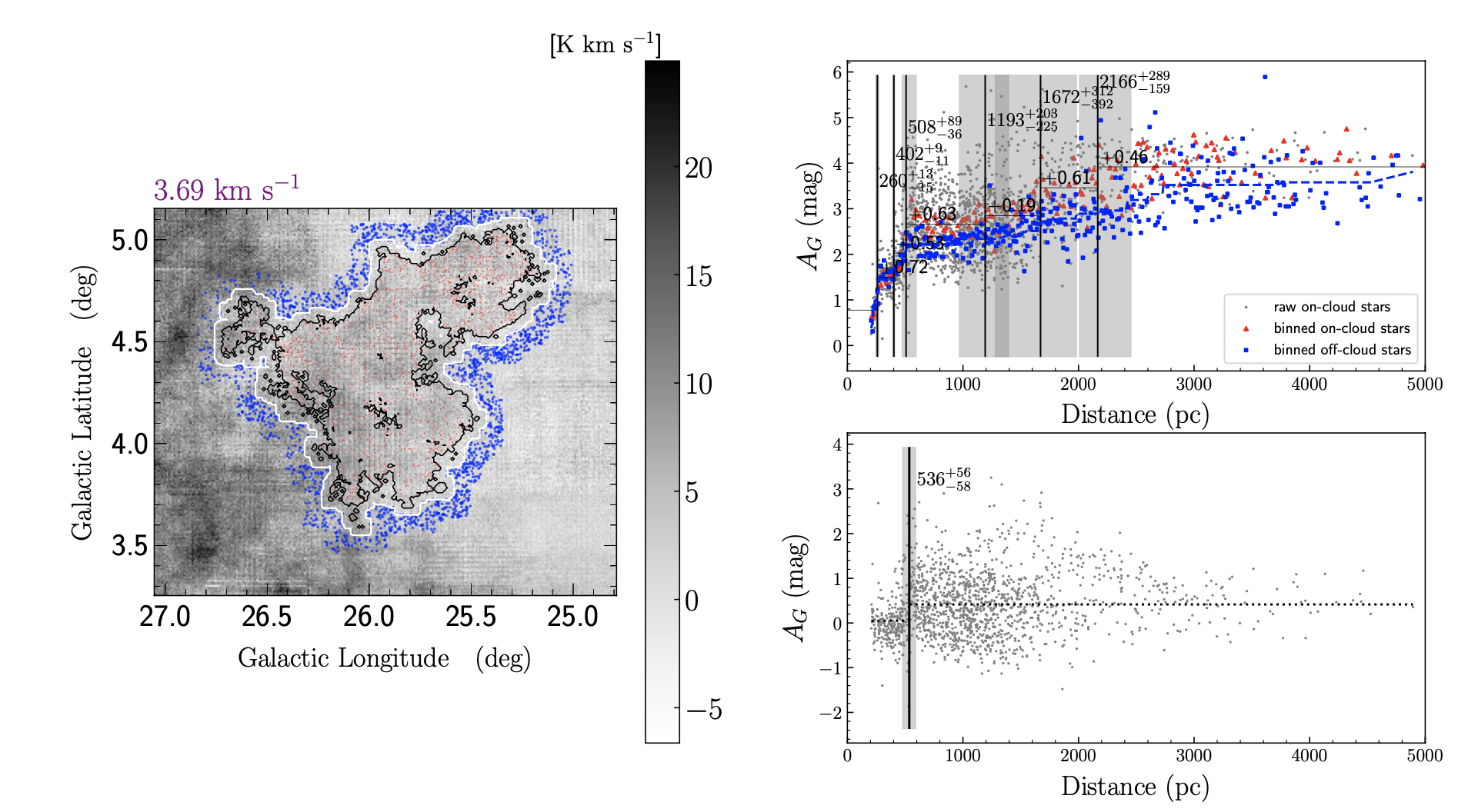}
\caption{ 
Distance measurement of a $^{13}$CO cloud structure toward the Serpens region using CU-JADE and BEEP. 
\label{fig:fig2_2}}
\end{figure*}

Our exclusion strategy depends on the reliability of the following three aspects:
\begin{enumerate}
    \item \textit{Robustness of field star selection and background subtraction:} By restricting the sample to stars with relative parallax uncertainties ($\sigma_\pi / \pi < 0.2$), we select field stars in ($2'$, $8'$) away from the cloud boundary \citep[more details see the Model B in][]{2024AJ....167..220Z}. For clouds with anomalous computations but plausible manual distance estimates, we manually refined the outer boundary ranges. The sub-arcminute CO data optimize cloud-mask definition for on/off-cloud star selection: at this resolution, mask pixels and stars are comparable in number ($\sim1.5$ stars per half-arcminute pixel within 5 kpc), making stellar sample selection sensitive to mask variations. Moreover, the porous ISM requires small pixels to properly capture true extinction variations along each LOS.

    \item \textit{Sensitivity to weak extinction components:} Based on existing simulations and observations \citep{2007A&A...470..615P, 2018MNRAS.474.4672L, 2023ApJ...943..172L}, CO typically forms in regions with $\Delta A_{V}\gtrsim$ 0.2--0.4 mag. The CU-JADE algorithm detects jumps as small as $\Delta A_G \lesssim 0.2$ mag, and enables tuning of the sensitivity to the stellar density, especially for the near and farthest side along LOS. These sensitivities prevent the misclassification of strong extinction features as the target cloud distance due to the omission of weaker, nearer jumps.
    
    \item \textit{Error propagation and handling:} For each extinction jump, we estimate the $1\sigma$ uncertainty via bootstrap resampling \citep[detailed in][] {2025AJ....170..185Z}. If the  distances of multiple extinction jumps fall in the uncertainties of the target jump (after BEEP; lower right panel in Figures~\ref{fig:fig2} and \ref{fig:fig2_2}), the final distance is generally assigned to the jump with the strongest $\Delta A_G$. In most cases, only the nearest jump falls within the 1$\sigma$ uncertainty range (see following examples). 
\end{enumerate}

We illustrate the necessity of this cleaning procedure using a large cloud L977 (UID: 3928144; Figures~\ref{fig:fig2}, \ref{fig:fig6}) toward Cyg OB7. We resolve at least four distinct CO components: Component 1 ($\rm 673^{+81}_{-73}\ pc$) at $v_{\mathrm{LSR}} \sim 0\ \mathrm{km\,s^{-1}}$, Component 2 ($\rm 579^{+15}_{-14}\ pc$) at $v_{\mathrm{LSR}} \sim 1\ \mathrm{km\,s^{-1}}$, a foreground layer with patched clouds at $v_{\mathrm{LSR}} \sim 5\ \mathrm{km\,s^{-1}}$ and $D \sim 300$--$400$ pc (Figure~\ref{fig:fig6}), and a more distant component ($v_{\mathrm{LSR}} <-5\ \mathrm{km\,s^{-1}}$) likely belonging to the inter-arm region (Figures~\ref{fig:fig2} and \ref{fig:fig8}). After applying BEEP background subtraction, the extinction jump corresponding to Component 1 becomes clear at $D \sim 670$ pc (see red triangles in Figure \ref{fig:fig2}). The method avoids erroneously attributing the cloud's distance to Component 2 ($\sim 550$ pc) or other more distant clouds, despite the presence of these significant jumps at various distances. Alternatively, single-jump method would simply converge to a mean value between these two closely spaced jumps (see Appendix \ref{appa}, after BEEP but a single-jump model). CU-JADE detects both jumps and associates them with their respective velocity components. 

Such contamination scenarios are prevalent throughout the inner Galaxy, and our methodology successfully removes the vast majority of these false associations combining CU-JADE and BEEP.
For example, an inner-Galaxy cloud (UID 3673804; see Figures~\ref{fig:fig2_2}, \ref{fig:fig6}, and \ref{fig:fig11}) is located near the junction of the Serpens Main and Serpens Near (see Section \ref{v1}) regions along the Galactic plane. At least two significant extinction jumps are evident within 1 kpc (top-right panel of Figure~\ref{fig:fig2_2}), caused by dust in the Serpens Main ($D \sim 508$ pc; $\Delta A_G \sim 0.63$ mag) and Serpens Near ($D \sim 260$ pc; $\Delta A_G \sim 0.72$ mag), respectively. 

By analyzing the extinction of on-cloud stars versus field stars, we determine that the two groups begin to diverge at $D = 508^{+89}_{-36}$ pc (see top-right panel in Figure \ref{fig:fig2_2}). This divergence confirms that the additional extinction of the on-cloud stars originates from the target cloud, despite the presence of stronger foreground extinction at 260 pc. After subtracting the background contribution using BEEP, CU-JADE successfully detects a single, prominent jump at this distance. In cases where multiple weak jumps persist after background subtraction, which is potentially caused by multiple similar dust structures along the LOS or the intrinsic complexity of the gas--dust geometry, we retain these candidates for subsequent analysis elsewhere. Nevertheless, the jump corresponding to the target cloud remains the most statistically significant in terms of $\Delta A_G$ and distance uncertainty $\sigma_d$.

\begin{figure*}[ht!]
\plotone{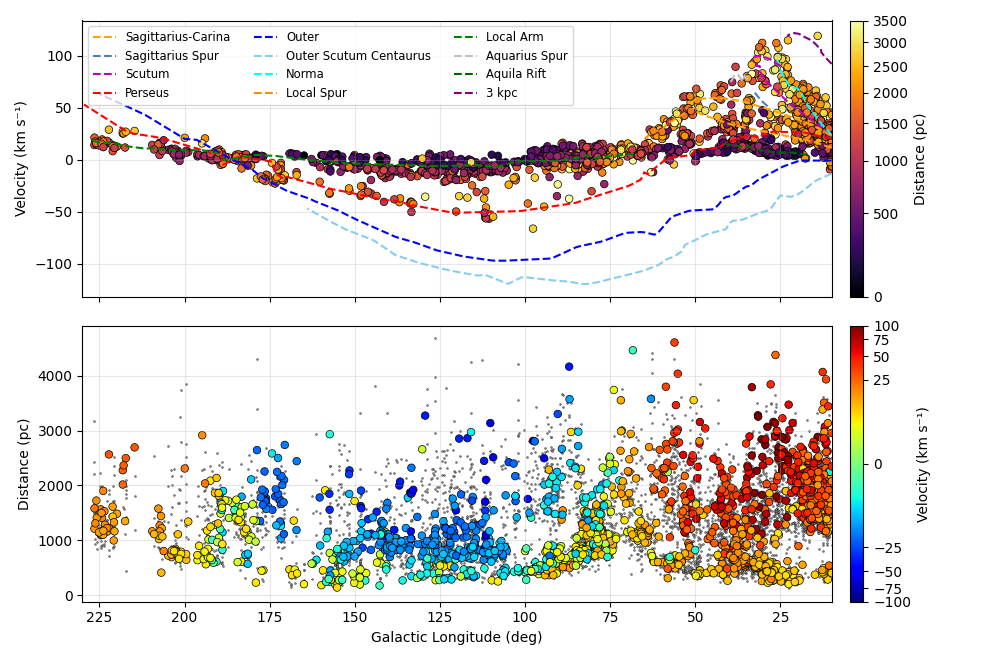}
\caption{Distributions of the 1,751 $^{13}$CO cloud sample in the $l-v$ and $l-d$ spaces. 
\textit{Top}: Centroid velocity of the clouds as a function of Galactic longitude. Colors indicate the distance, and the dashed colored lines represent the spiral arm model from \citet{2014ApJ...783..130R}. \textit{Bottom}: Distance of the clouds as a function of longitude (error bars omitted). Colors indicate the centroid velocity. The gray dots represent additional extinction jump points detected along the same LOS, suggesting the presence of other strong extinction components (foreground or background clouds). The color scale is non-linear to emphasize the large population of cloud structures at low velocities and nearby distances.
\label{fig:fig3}}
\end{figure*}

\section{cloud catalog}
\label{catalog}

Ultimately, using the methodology described in Section \ref{tech}, we performed distance measurements for approximately 3,600 cloud structures with angular areas larger than $\sim 43.25$ square arcminutes. Initial distance solutions were obtained for 1,969 of these clouds. The 1,631 cloud structures lack distance measurements primarily due to the absence of jump detection results after BEEP processing. Through manual inspection, combined with analyses of the linewidth-size relation, molecular gas scale heights, and the distance-velocity relation, we classified 218 clouds (Class III) as suffering from foreground contamination (see Appendix \ref{appc}). For these clouds, the distance estimates are likely underestimated due to the presence of numerous clumpy foreground structures that cannot be effectively disentangled from the target cloud. These contaminated clouds are primarily distributed toward the Aquila Rift in the inner Galaxy and are excluded from the subsequent analysis. 

The remaining 1,751 MCs are considered to have robust distance measurements (Table \ref{table:tab1}). From a statistical robustness perspective, we further classified 170 clouds as Class II, where the extinction jump at the estimated distance is less than 0.4 mag or the number of on-cloud stars is fewer than 30. The rest are designated as Class I. The velocity and distance distributions of these clouds (class I+II) are presented in Figure~\ref{fig:fig3}. Table~\ref{table:tab1} contains the basic information and physical properties. Columns such as the extinction increment before and after the BEEP fitting, along with other intermediate diagnostic values, will appear in two separate full catalogs \citep{scidbv2}  \footnote{https://doi.org/10.57760/sciencedb.0121a}. In particular, the post-BEEP extinction increment and its uncertainty can serve as reliability metrics for individual cloud distance measurements \citep[e.g., in][we adopted a threshold of $\Delta A_G \gtrsim$ 0.15~mag]{2024AJ....167..220Z}.  

Based on Figure~\ref{fig:fig3}, we find that MCs within Solar neighbourhood broadly follow the $l$–$v$–$d$ relation predicted by the Galactic rotation and \citet{2014ApJ...783..130R} spiral arm models. However, significant deviations in velocity and distance are evident for many clouds, particularly those in the Perseus Arm of the second quadrant. These offsets are likely driven by local dynamical processes, manifesting as the Burton's Curse effect \citep{1971A&A....10...76B, 2022ApJ...925..201P}.

\begin{longrotatetable}
\begin{deluxetable*}{ccccccccccccc}
\setlength{\tabcolsep}{3.2mm}
\tablecaption{Parameters of identified molecular clouds\label{chartable}}
\tablewidth{720pt}
\tabletypesize{\scriptsize}
\tablehead{
\colhead{(1)} & \colhead{(2)} & \colhead{(3)} & \colhead{(4)} & \colhead{(5)} &
\colhead{(6)} & \colhead{(7)} & \colhead{(8)} & \colhead{(9)} &
\colhead{(10)} & \colhead{(11)} & \colhead{(12)} & \colhead{(13)} \\
\colhead{id} & \colhead{uid} & \colhead{$l$} & \colhead{$b$} & 
\colhead{$v_{\rm LSR}$} & \colhead{$\sigma_{v}$} & \colhead{$R$} & 
\colhead{$\mathrm{Distance}$} & 
\colhead{$N_{\rm mean}$} & \colhead{$Mass^{\rm LTE}$} &
\colhead{$Mass^{\rm Xfactor}$} &
\colhead{$\alpha$} &
\colhead{Class} \\ 
\colhead{} & \colhead{} & \colhead{deg} & \colhead{deg} & \colhead{$\mathrm{km\cdot s^{-1}}$} & \colhead{$\mathrm{km\cdot s^{-1}}$} &
\colhead{pc} & 
\colhead{pc} & \colhead{$\mathrm{cm^{-2}}$} &
\colhead{$m_{\odot}$} & \colhead{$m_{\odot}$} &\colhead{ } & 
\colhead{ } \\
}
\startdata
1  &  3014957 &   9.80 &   1.27 &   13.4 &    1.4 &   3.6 &  $1793^{+281}_{-215}$ &  2.37e+21 &   2020 &          4540 &         4.3 &     i \\
2  &  2804868 &   9.80 &   1.73 &   12.3 &    1.3 &   3.6 &  $1751^{+301}_{-274}$ &  2.26e+21 &   1950 &          4180 &         3.9 &     i \\
3  &  2000602 &   9.88 &  -0.14 &   42.9 &    1.0 &   3.9 &  $2348^{+404}_{-321}$ &  4.85e+21 &   4890 &          4980 &         0.9 &    ii \\
4  &  4053204 &   9.90 &  -0.50 &   -6.4 &    3.3 &   3.6 &  $1679^{+112}_{-152}$ &  4.11e+21 &   3540 &          6910 &        13.1 &     i \\
5  &  2600034 &   9.91 &  -0.56 &   28.8 &    2.5 &   6.2 &  $1887^{+301}_{-298}$ &  9.59e+21 &  24310 &         33130 &         1.9 &     i \\
6  &  3476195 &   9.96 &  -0.03 &   11.2 &    2.5 &   2.1 &  $1507^{+166}_{-185}$ &  5.56e+21 &   1670 &          3720 &         9.4 &     i \\
7  &  2884658 &   9.96 &   3.42 &    2.6 &    0.6 &   2.0 &    $1384^{+79}_{-57}$ &  9.70e+20 &    250 &           420 &         3.5 &     i \\
8  &  3850247 &   9.96 &  -0.08 &   17.6 &    2.4 &   7.1 &  $2076^{+346}_{-426}$ &  4.86e+21 &  16190 &         28800 &         2.9 &     i \\
9  &  4160869 &   9.97 &  -1.03 &   34.7 &    2.8 &   4.4 &  $2960^{+378}_{-418}$ &  5.67e+21 &   7380 &         15920 &         5.2 &     i \\
$\vdots$ & $\vdots$ & $\vdots$ & $\vdots$ & $\vdots$ & $\vdots$ & $\vdots$ & $\vdots$ & $\vdots$ & $\vdots$ & $\vdots$ & $\vdots$ & $\vdots$ \\
1743 &  1808274 &  224.67 &  -1.05 &   17.7 &    0.6 &   1.9 &  $1565^{+270}_{-312}$ &  2.78e+21 &    660 &           730 &         1.2 &     i \\
1744 &  2433822 &  224.74 &  -1.76 &   16.2 &    1.1 &   2.0 &    $1128^{+59}_{-64}$ &  4.37e+21 &   1160 &          1420 &         2.4 &     i \\
1745 &  2824948 &  224.91 &   0.74 &   17.3 &    0.5 &   2.2 &  $1487^{+195}_{-214}$ &  1.16e+21 &    390 &           530 &         1.7 &     i \\
1746 &  3577542 &  225.14 &  -1.10 &   13.6 &    0.9 &   2.3 &    $1144^{+61}_{-31}$ &  1.91e+21 &    670 &           760 &         3.1 &     i \\
1747 &  1469190 &  225.81 &  -0.84 &   14.5 &    0.5 &   2.1 &    $1412^{+98}_{-74}$ &  1.66e+21 &    510 &           730 &         1.5 &     i \\
1748 &  3693875 &  225.92 &  -0.40 &   15.9 &    1.0 &  10.1 &  $1725^{+326}_{-275}$ &  2.23e+21 &  15070 &         18450 &         0.7 &     i \\
1749 &  3463750 &  226.20 &  -0.62 &   17.4 &    0.9 &   3.1 &  $1310^{+203}_{-123}$ &  1.98e+21 &   1300 &          1570 &         2.5 &     i \\
1750 &  2354054 &  226.47 &   1.39 &   21.2 &    0.4 &   3.8 &  $1584^{+155}_{-149}$ &  1.10e+21 &   1040 &          1170 &         0.7 &     i \\
1751 &  1702558 &  226.48 &  -2.37 &   14.0 &    0.7 &   1.4 &    $1213^{+33}_{-25}$ &  1.71e+21 &    220 &           270 &         3.4 &     i \\
\enddata 
\tablecomments{
(1) id: Cloud ID in this work of distanced measured MCs, sorted by longitude.\\
(2) uid: The unique ID of each MC, facilitate fast browser of the target MC. \\
(3) $l$: Longitude of cloud center weighted by intensity.\\
(4) $b$: Latitude of cloud center weighted by intensity.\\
(5) $v_{\rm LSR}$: Velocity of cloud center weighted by intensity.\\
(6) $\sigma_{v}$: Intensity weighted velocity dispersion of clouds.\\
(7) R: Physical size of molecular cloud, in unit of pc.\\
(8) $\rm Distance$: Median cloud distance derived from the bootstrap procedure. The uncertainties correspond to the 16th and 84th percentiles of the cloud distance distribution obtained from 10,000 resampled jump measurements (Lower/Upper value of $1\sigma$ confidence interval).\\
(9) $N_{\rm mean}$: The mean column density of each cloud based on LTE. \\ 
(10) $Mass^{\rm LTE}$: LTE mass derived from $\rm ^{13} CO$ emission with determined distance, which is calculated by column density of $\rm ^{13} CO$ emission.\\
(11) $Mass^{\rm Xfactor}$: Mass calculated by $\rm ^{12} CO$ emission with determined distance, in a $\rm ^{13} CO$ traced cloud boundary. The CO-to-$\rm H_{2}$ Conversion Factor adopts $\rm X_{CO} = 2 \times 10^{20} cm^{-2} (K\cdot km\cdot s^{-1})^{-1}$. \\
(12) Virial parameter of $\rm ^{13}CO$ traced cloud. \\ 
(13) Class: The quality flag of distance measured MCs. \\
Other quantities along with distance measurement can be seen in full machine-readable format in \href{10.57760}{https://doi.org/10.57760/sciencedb.0121a}. 
}
\end{deluxetable*}
\end{longrotatetable}%
\label{table:tab1}
\vspace*{-\baselineskip}

\subsection{Statistics and Completeness of Cloud Sample \label{stat}}
The completeness of MC distance measurements is difficult to quantify as a function of heliocentric distance. We therefore adopt the 3,600 clouds of larger angular size as a parent sample. Both this sample and the full catalog of 30,626 clouds serve as denominators for number counts and flux fractions, providing an indirect completeness assessment regarding the distance measurement success rate and angular-area selection. We note that the current latitude coverage of MWISP I does not include some high-latitude clouds in the solar vicinity. Since such clouds are typically closer and more extended, they are more readily measured and have been systematically investigated in the literature \citep{2019ApJ...879..125Z, 2019A&A...624A...6Y}.  

\begin{figure*}[ht!]
\plotone{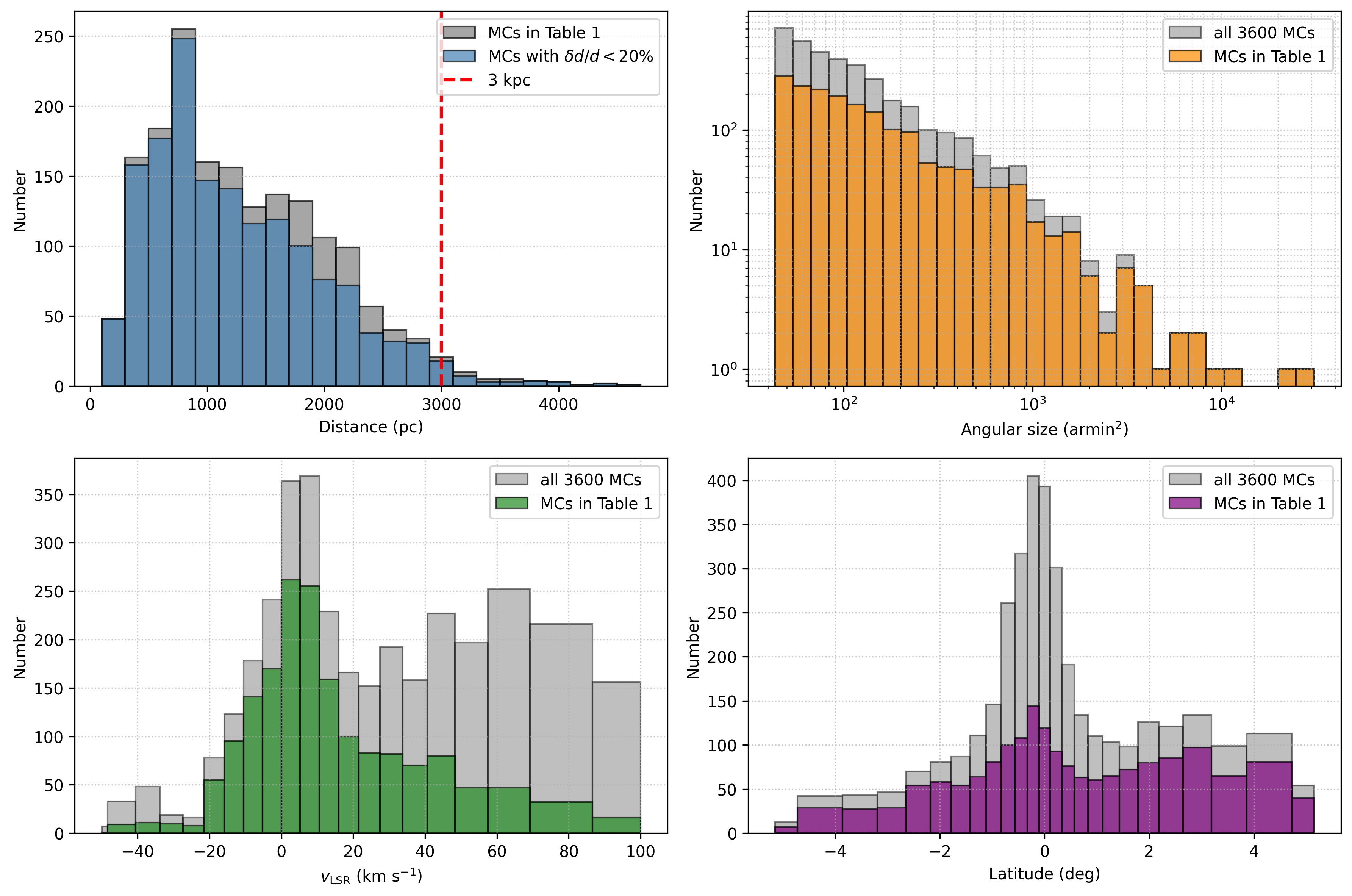}
\caption{Statistical comparison of the 1,751 MCs with successful distance measurements (listed in Table~\ref{table:tab1}) and the parent sample of 3,600 clouds with angular areas larger than $\sim 43.25~{\rm arcmin^{2}}$. These panels illustrate the completeness of the distance measurements under various conditions.
\textit{Top left:} Histograms of cloud counts versus distance. The gray histogram represents the 1,751 clouds in Table \ref{table:tab1}, while the blue histogram shows the distribution of clouds with distance uncertainties smaller than 20\%.
\textit{Top right:} Histograms of cloud counts versus angular area. The gray histogram represents the parent sample of 3,600 clouds, while the orange histogram shows the distribution of the 1,751 clouds with measured distances. 
\textit{Bottom left:} Histograms of cloud counts versus radial velocity ($v_{\text{LSR}}$). The gray histogram represents the parent sample of 3,600 clouds, while the orange histogram shows the distribution of the 1,751 clouds with measured distances. Non-uniform bins are adopted to better display the large number of clouds near 0 km s$^{-1}$.
\textit{Bottom right:} Histograms of cloud counts versus Galactic latitude ($b$). The gray histogram represents the parent sample of 3,600 clouds, while the orange histogram shows the distribution of the 1,751 clouds with measured distances. Non-uniform bins are adopted to better display the concentration of clouds near the Galactic plane. }
\label{fig:fig5}
\end{figure*}

\begin{table}[htbp]
\centering
\caption{Statistics of cloud catalogues: cloud counts and integrated flux.}
\label{tab:flux_stats}
\setlength{\tabcolsep}{24pt}
\begin{tabular}{lccccl}
\toprule
 & \multicolumn{2}{c}{Cloud Number} & \multicolumn{2}{c}{Flux$^a$} & Notes \\
\cmidrule(lr){4-5}
 & & & \multicolumn{2}{c}{(K km s$^{-1}$ arcmin$^2$)} & \\
\midrule
\multirow{2}{*}{Full Cloud Catalog} & \multirow{2}{*}{30,626} & 21,614 & \multirow{2}{*}{5,302,985} & 4,320,710 & inner \\
 & & 9,012 & & 982,275 & outer \\
\midrule
\multirow{2}{*}{Parent Sample} & \multirow{2}{*}{3,600} & 2,625 & \multirow{2}{*}{4,195,329} & 3,406,874 & inner \\
 & & 975  & & 788,455 & outer \\
\midrule
\midrule
\multirow{2}{*}{Table \ref{table:tab1}} & \multirow{2}{*}{1,751} & 1,086  & \multirow{2}{*}{2,241,283} & 1,612,184 & inner \\
 & & 665  & & 629,099 & outer \\
\midrule
\bottomrule
\end{tabular}

\medskip
\footnotesize
$^a$ Fluxes are derived by integrating all Gaussian components. 
\end{table}

The overall success rate for distance measurements in the parent sample (i.e., the 3,600 targeted clouds) is 48.64\%, although these clouds account for only 5.72\% of the total cloud catalog. This discrepancy arises because the majority of clouds are distributed at small angular areas, roughly following an exponential distribution \citep[Figure~\ref{fig:fig5}; see also][]{2020ApJ...898...80Y}. These small clouds lack a sufficient number of on-cloud stars, leading to failed or unreliable distance measurements, as evidenced by the exponentially rising failure rate with decreasing angular area shown in Figure~\ref{fig:fig5}. Consequently, clouds with angular areas above $\sim 50$ square arcminutes \citep[e.g.,][empirical cloud scale for Gaia DR3 stellar number density]{2024AJ....167..220Z} are considered suitable for distance estimation. For some clouds at the small end of the angular area distribution, we extrapolate the cloud boundary by 0.5--1.5 arcmin to include more on‑cloud stars, which yielded more reliable distance estimates when the dust extinction extends beyond the $^{13}$CO emission.

The 1,751 MCs with measured distances (Table~\ref{table:tab1}) account for 53.42\% of the total flux of the parent sample and 42.26\% of the total flux of the full cloud catalog, i.e., the fraction of the cataloged $^{13}$CO flux contained in clouds with accepted distances. This indicates that the small-scale clouds omitted during distance measurement contribute only a minor fraction weighted by flux. Moreover, the success rate remains substantial for many giant MCs located at large distances, low Galactic latitudes, and high radial velocities, validating the effectiveness of the distance measurement methodology described in Section \ref{tech}.

Top left panel in Figure~\ref{fig:fig5} implies observational bias at different distances. Beyond 2 kpc, and particularly beyond 3 kpc, the number of clouds with measured distances drops sharply. This decline is primarily due to the limitations of \textit{Gaia} optical observations at large distances, coupled with increasing distance uncertainties. This also reflects the fact that distant clouds with large angular scales are naturally rarer due to distance effects. Similarly, for clouds with high radial velocities ($v > 50$ km s$^{-1}$), the success rate falls below 30\%. This suggests that high-velocity clouds are typically located at greater distances, consistent with predictions from Galactic rotation curves (as shown in Figure~\ref{fig:fig3}).

Clouds at low Galactic latitudes ($|b| < 1^{\circ}$) exhibit a similar trend. They are likely distant structures whose apparent positions are projected onto the Galactic plane. Additionally, lower latitudes correspond to higher LOS crowding and contamination, further reducing the success rate. 
We further note that clouds located in the inner-Galaxy ($l < 90^{\circ}$) presented in Table~\ref{table:tab1} contribute approximately 47.32\% of the total flux from the inner-Galaxy parent sample, and those in the outer-Galaxy
($l > 90^{\circ}$) contribute 79.79\%. The corresponding flux contributions
to the full catalog are 37.31\% (inner) and 64.05\% (outer). This indicates a lower distance measurement success rate in the inner-Galaxy compared to the second and third quadrants, a discrepancy driven by the prevalence of more distant clouds with higher extinction in the inner disk region. Nevertheless, we conclude that, in terms of the flux, nearly half of the molecular gas in the parent sample yields distance estimates. Counts and flux statistics for the distance-accepted sample (Table \ref{table:tab1}), the parent sample, and the full cloud catalog are given in Table \ref{tab:flux_stats}.

\begin{figure*}[ht!]
\plotone{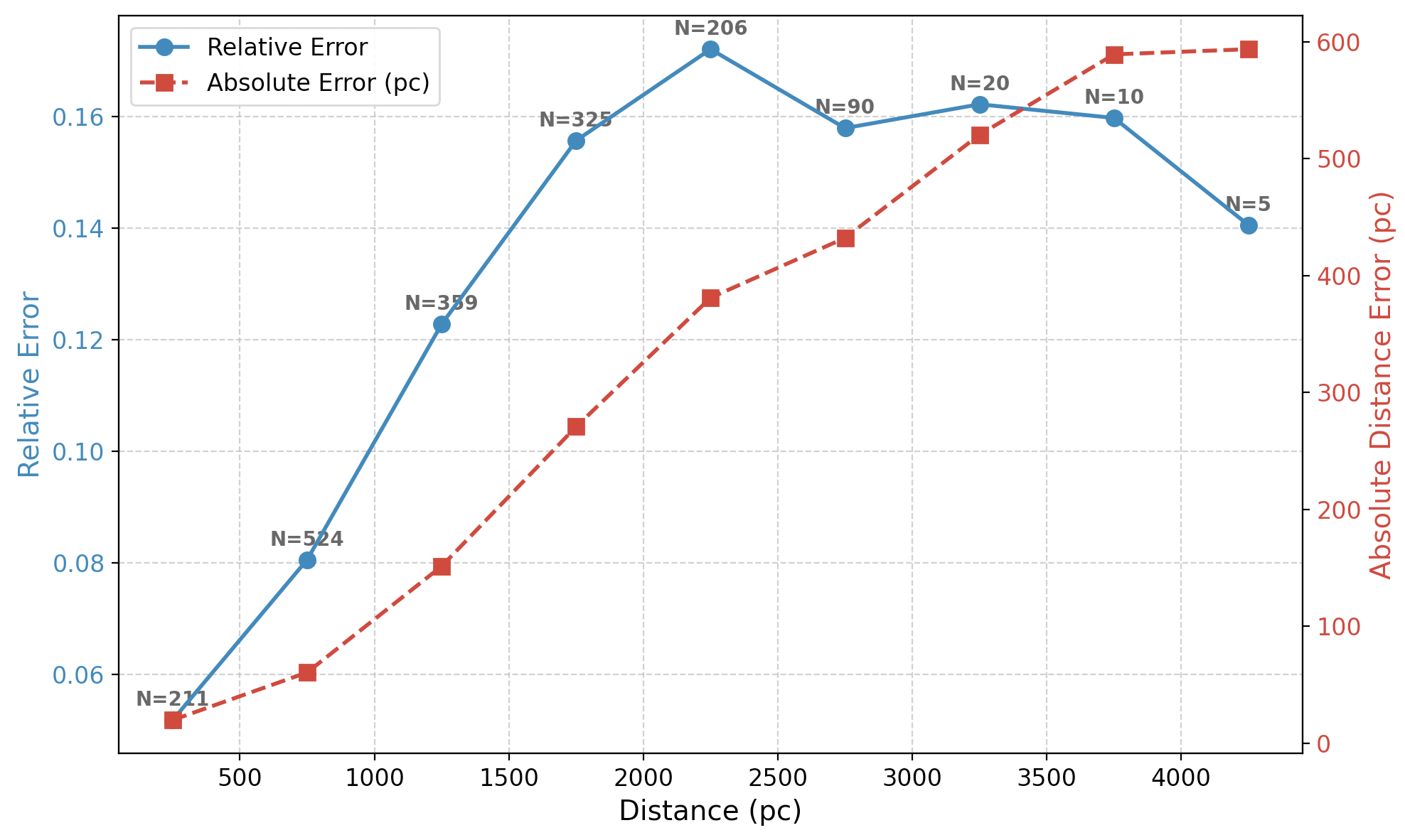}
\caption{Distance uncertainties of the cloud sample as a function of heliocentric distance. Blue and red lines denote the relative and absolute distance errors, respectively. The values are averaged within 500 pc bins, and the number of clouds in each bin is indicated at the corresponding positions. }
\label{fig:fig45}
\end{figure*}

\subsection{Uncertainty of Distance Measurements \label{err}} 
We quantitatively assess the distance uncertainties of the cloud sample (Table \ref{table:tab1}) and their potential impact on subsequent analyses. For the absolute distance errors, we obtain a symmetric error value by taking the difference between the 84th and 16th percentiles of cloud distances and dividing by two. The sample is binned by 500 pc intervals to trace the variation in distance uncertainty with heliocentric distance. As illustrated by the red and blue lines in Figure \ref{fig:fig45}, the statistical distance uncertainty increases monotonically up to 2 kpc, with absolute errors continuing to rise at larger distances.

Within 1 kpc, using the stated symmetric-error convention, the mean relative distance error is below 10\%, corresponding to an absolute uncertainty of $\lesssim 100$ pc. The cloud count increases nearly threefold, from 211 within 500 pc to 735 within 1 kpc. This implies that, when larger-scale structures are adopted as the parent sample, the number recovery fraction of cloud distance measurements within 1 kpc is empirically higher than that beyond 1 kpc. Consequently, this subset constitutes a ``gold sample" in terms of both accuracy and recovery fractions (see Section \ref{bird}).

In the 1--2 kpc bin, the average relative error rises to 13.8\%, with an absolute uncertainty of 209 pc. While this level of uncertainty begins to affect the identification of coherent structures manifesting as LOS stretching, it remains lower than that of more distant clouds (represented by red squares in Figure \ref{fig:fig45}). Furthermore, despite significant incompleteness, the cloud count in this bin does not exhibit a sharp decline, leaving this bin well populated and still contributing substantially to the catalog. These statistically robust results provide the necessary basis for classifying the spatial distribution of molecular gas in Section \ref{bird}.

\subsection{Accuracy of Distance Measurements \label{accuracy}}

Maser emission is generally associated with circumstellar envelopes or outflows of massive stars. Trigonometric parallax measurements obtained from Very Long Baseline Interferometry (VLBI) arrays are regarded as the most accurate geometric distance determinations within the Milky Way and beyond. We compiled the 199 maser sources with distance measurements from \citet{2019ApJ...885..131R} and performed a cross-match with the clouds listed in Table~\ref{table:tab1}.

\begin{figure*}[ht!]
\plotone{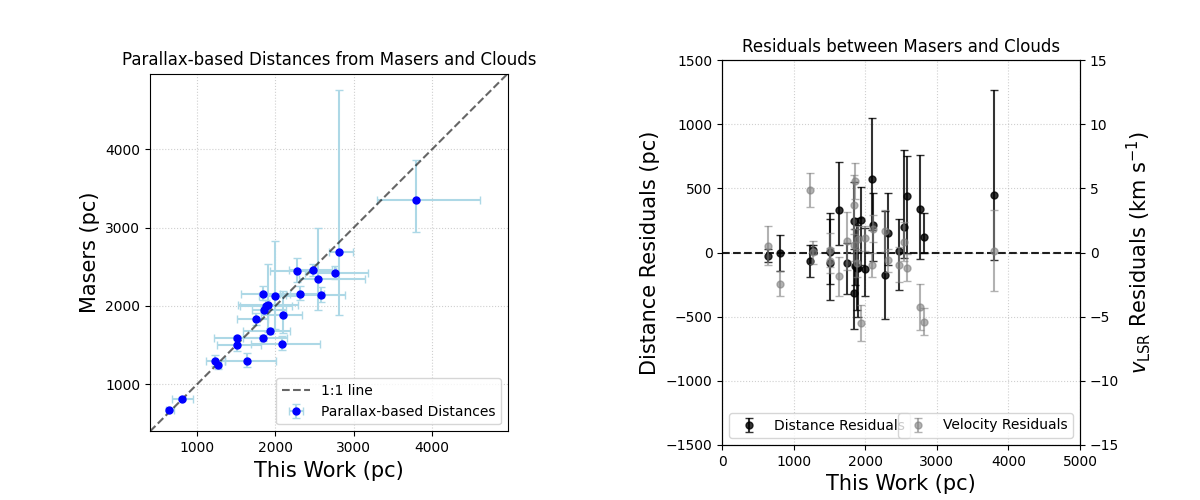}
\caption{Comparison of 25 retained maser–cloud pairs involving 24 masers. Pairs satisfy spatial overlap and a $15~\mathrm{km~s^{-1}}$ velocity cut, with G192.16-03.81 included manually. The two G059.78+00.06 pairs share one parallax, and distance agreement was one factor in choosing between the G188.79+01.03 candidates. 
\textit{Left panel:} Comparison between maser distances derived from trigonometric parallax measurements and MC distances obtained via extinction jumps. The two independent distance measurements agree well within their respective uncertainties.
\textit{Right panel:} Black points are residuals between MC distances in this work and parallax-based distances of masers. The vertical error bars represent the distance uncertainties from Table~\ref{table:tab1}. Greys points are residuals between the centroid velocities of these MCs and the maser radial velocities reported in \citet{2019ApJ...885..131R}. The vertical error bars indicate the velocity dispersion of the MCs in Table~\ref{table:tab1}.} 
\label{fig:fig4}
\end{figure*}

Among these 199 masers, 168 fall within the projected data coverage, and 81 in 168 are within 4~kpc. However, only 40 masers have coordinates that overlap with the cloud structures in Table~\ref{table:tab1}. We used the maser velocity information for further filtering: if the velocity difference exceeds $15~\mathrm{km~s^{-1}}$, the maser is rejected, leaving only 26 sources that are matched in both spatial projection and velocity. In addition, during manual inspection, we found one source, G192.16-03.81, whose maser velocity is consistent with the $^{13}$CO centroid velocity but is slightly offset spatially. Given that we adopted a high intensity threshold of 5$\sigma$, we also included it in the matching. This resulted in 27 MCs being spatially associated with the HMSFRs traced by the masers. 

These 27 cloud--maser pairs involve 25 distinct maser sources, meaning that two masers each have two different matching clouds. The two masers are G059.78+00.06 and G188.79+01.03. For the latter, UID 3649271 shows larger discrepancies in both velocity and distance, and we exclude it from the matched sample. For G059.78+00.06, we could not reliably determine which of the two matched clouds is associated with the maser, and thus both are retained in this work. Additionally, G017.55--00.12 \citep[IRC--10414;][]{2014MNRAS.437..843G} is a runaway red supergiant (RSG) hosting maser emission in its circumstellar envelope, and as a precaution, we exclude the M7-type RSG IRC--10414 from the statistical analysis of CO cloud--maser distances. Ultimately, we obtain 25 matched pairs of maser sources and $^{13}$CO cloud structures. As shown in Figure~\ref{fig:fig4}, the distances and radial velocities of these matched sources exhibit excellent agreement within their respective uncertainties, with no apparent systematic offset.

Notably, 14 of the 25 matched clouds are located in the inner Galaxy (first quadrant). These results demonstrate that our methodology is both feasible and effective even in the complex environments of the inner Galactic plane. As shown in Section \ref{bird}, the face-on view of the Solar System shows that these masers exhibit good spatial correspondence with the high-density regions traced by $^{13}$CO MCs, particularly for those within 3 kpc.

\subsection{Physical Property Calculations \label{pp}}

Following the methodology of \citet{2024AJ....167..220Z}, we provide only a brief overview of the calculation here. We first compute the center coordinates ($l, b$), centroid velocity ($v_{\text{LSR}}$), and velocity dispersion ($\sigma_v$) using flux-weighted averages. The angular size of each cloud is then derived from the number of pixels within its boundary. Under the assumption of LTE, the column density is calculated from the $^{13}$CO integrated intensity, the excitation temperature ($T_{\text{ex}}$) is determined from the optically thick $^{12}$CO ($J=1\text{--0}$) emission. Finally, combining these with the measured distances, we derive the effective radius ($R$), mass ($M$), and virial parameter ($\alpha$). The mass traced by $^{12}$CO is independently estimated from the integrated $^{12}$CO intensity within the PPV boundaries, adopting the standard conversion factor $X = 2 \times 10^{20}$ cm$^{-2}$ (K km s$^{-1}$)$^{-1}$ \citep{2013ARA&A..51..207B}. All derived physical parameters are listed in Table~\ref{table:tab1}. A complete version of this table, including additional parameters, is available in the electronic edition online.

\section{The 3D distribution and velocity field of molecular clouds}
\label{sect5}

\subsection{Overview of MCs in various Velocities and Distances} 
\label{scroll}

\begin{figure*}[ht!]
\includegraphics[width=18cm,angle=0]{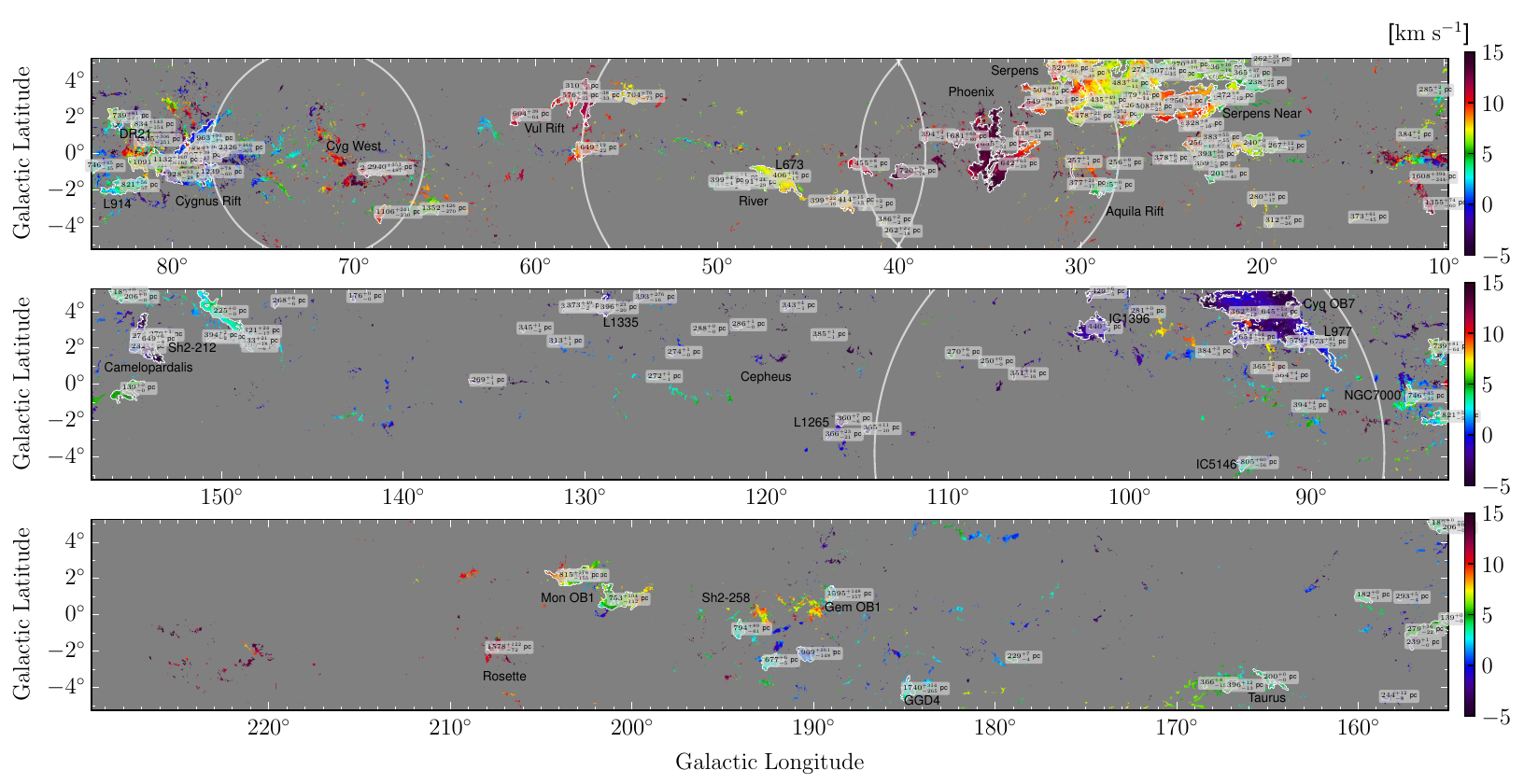}
\caption{Distribution of the $^{13}$CO centroid velocities. 
For display purposes, only low-velocity MCs are shown, defined as those with centroid velocities within the range of $(-5, 15)$ km s$^{-1}$. The color bar limits are correspondingly restricted to this velocity interval. Boundaries of nearby or large angular-size clouds (sub-sample) are outlined, with distances labeled in black text on a gray background. Several well-known regions along the northern Galactic plane, as cataloged in the Star Formation Handbook, are also labeled for reference. The white circles/arcs indicate the approximate projected extent of the large-scale bubble structures identified in the face-on map (Sections \ref{bird} and \ref{subregion}). This map can be compared with the MC distributions in other velocity intervals (see Figures~\ref{fig:fig7} and \ref{fig:fig8}).}
\label{fig:fig6}
\end{figure*}

Consistent with the data release paper and for the convenience of the reader, we present the $\sim 2300$ deg$^2$ ``scroll'' by unfolding it into three contiguous sections. To mitigate complexity along the LOS, the velocity range is divided into three intervals: the low-velocity interval ($-5$ to $15$ km s$^{-1}$; Figure~\ref{fig:fig6}), the high-positive-velocity interval ($>15$ km s$^{-1}$; Figure~\ref{fig:fig7}), and the high-negative-velocity interval ($<-5$ km s$^{-1}$; Figure~\ref{fig:fig8}). 

\subsubsection{The low-velocity interval MCs}\label{v1}

In Figure~\ref{fig:fig6}, extensive $^{13}$CO emission is observed as expected, where we highlight large-scale, nearby structures. The most prominent regions with increasing longitude are the Serpens \citep{2018A&A...620A..62G, 2019ApJ...878..111H}, Aquila Rift \citep[including the River and Phoenix clouds;][]{2020ApJ...893...91S}, Cygnus \citep{2020ApJS..248...15Z, 2024AJ....167..220Z}, Cyg OB7, Cepheus \citep{1997ApJS..110...21Y}, Camelopardalis \citep{2008hsf1.book..294S}, Gem OB1 \citep{2017ApJS..230....5W, 2018ApJS..235...15L}, and Mon OB1 region \citep{2024ApJ...966..202Z}. With the exception of Cygnus X (located tangentially) and Gem OB1 (in the anti-Galactic-center direction), which are relatively distant, the bulk of these structures lie within 1 kpc.

Several interesting regions are systematically resolved here for the first time. For instance, the Serpens regions exhibit highly extended, structured features even in $^{13}$CO emission, where the $^{13}$CO cloud structures tend to trace the main body of the cloud rather than clumpy or filamentary substructures. Between $l \sim 20^{\circ}$ and $33^{\circ}$, two layers of MCs with distinct distances distribution is evident for clouds with angular areas $A > 500$ arcmin$^2$. This distribution is segregated by region: the Serpens Near (west, $l \sim 20^{\circ}\text{--}28^{\circ}$) resides in the foreground ($D \sim 227\text{--}260$ pc), while Serpens Main (east, $l \sim 26^{\circ}\text{--}33^{\circ}$) is significantly more distant ($D \sim 436\text{--}550$ pc). These two complexes overlap spatially at $l \sim 26^{\circ}\text{--}28^{\circ}$ and share very similar velocities (see Figures~\ref{fig:fig6} and \ref{fig:fig11}). Additionally, smaller-scale structures are found at intermediate distances ($D \sim 300\text{--}400$ pc). 

From a macroscopic perspective, the Aquila-Serpens complex exhibits a distance gradient ($D \sim 200\text{--}700$ pc) and a velocity gradient ($v_{\rm LSR}\sim 4\text{--}14$ km s$^{-1}$) extending eastward to the Phoenix MC. Combined with the River complex at $l \sim 40^{\circ}\text{--}50^{\circ}$ ($D \sim 400$ pc), these clouds form a prominent kinematic structure (Section \ref{split}). Centered on the Phoenix cloud, there is a trend of decreasing distances and radial velocities toward the periphery (Figure \ref{fig:fig6}; also see the inverted V-shaped feature in the $l\text{-}v$ and $l\text{-}d$ diagrams at $l \sim 35^{\circ}$ in Figure~\ref{fig:fig3}). 

\begin{figure*}[ht!]
\includegraphics[width=18cm,angle=0]{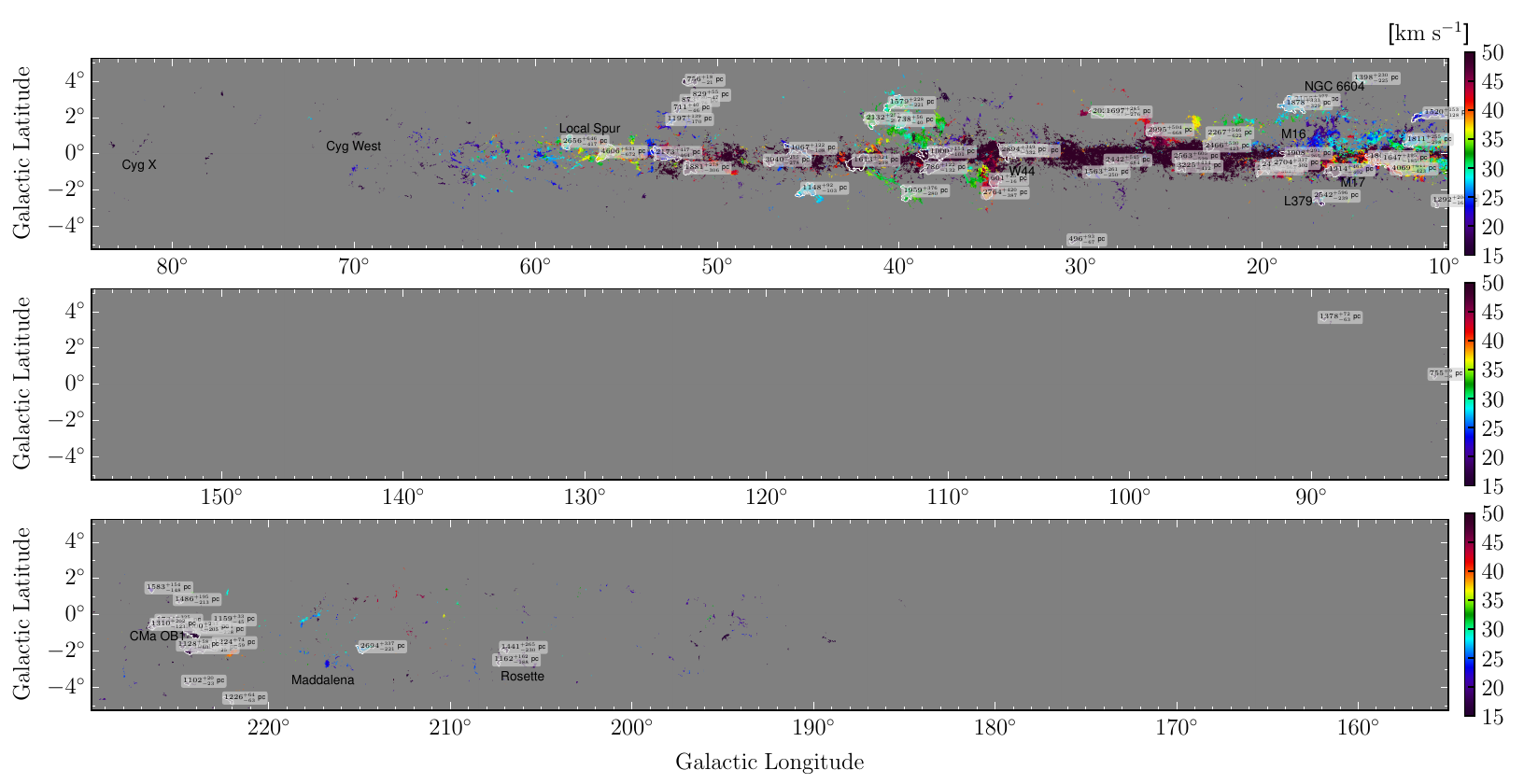}
\caption{Similar to Figure~\ref{fig:fig6}, but showing $^{13}$CO cloud structures with velocities greater than 15 km s$^{-1}$. To better display the results, lower-velocity components are plotted in the top, and the color bar is limited to the range of $(15, 50)$ km s$^{-1}$. Higher-velocity components are uniformly shown in dark red. Similarly, only part of cloud structures with large angular sizes are labeled. Contours outline the cloud boundaries, and distances are indicated in black text on a gray background. Prominent regions such as W44, M16, M17, and NGC 6604 possess accurately measured distances. Notably, and perhaps unsurprisingly, few $^{13}$CO cloud structures in the second quadrant appear in Table~\ref{table:tab1} at high-positive velocities. }
\label{fig:fig7}
\end{figure*}

\begin{figure*}[ht!]
\includegraphics[width=18cm,angle=0]{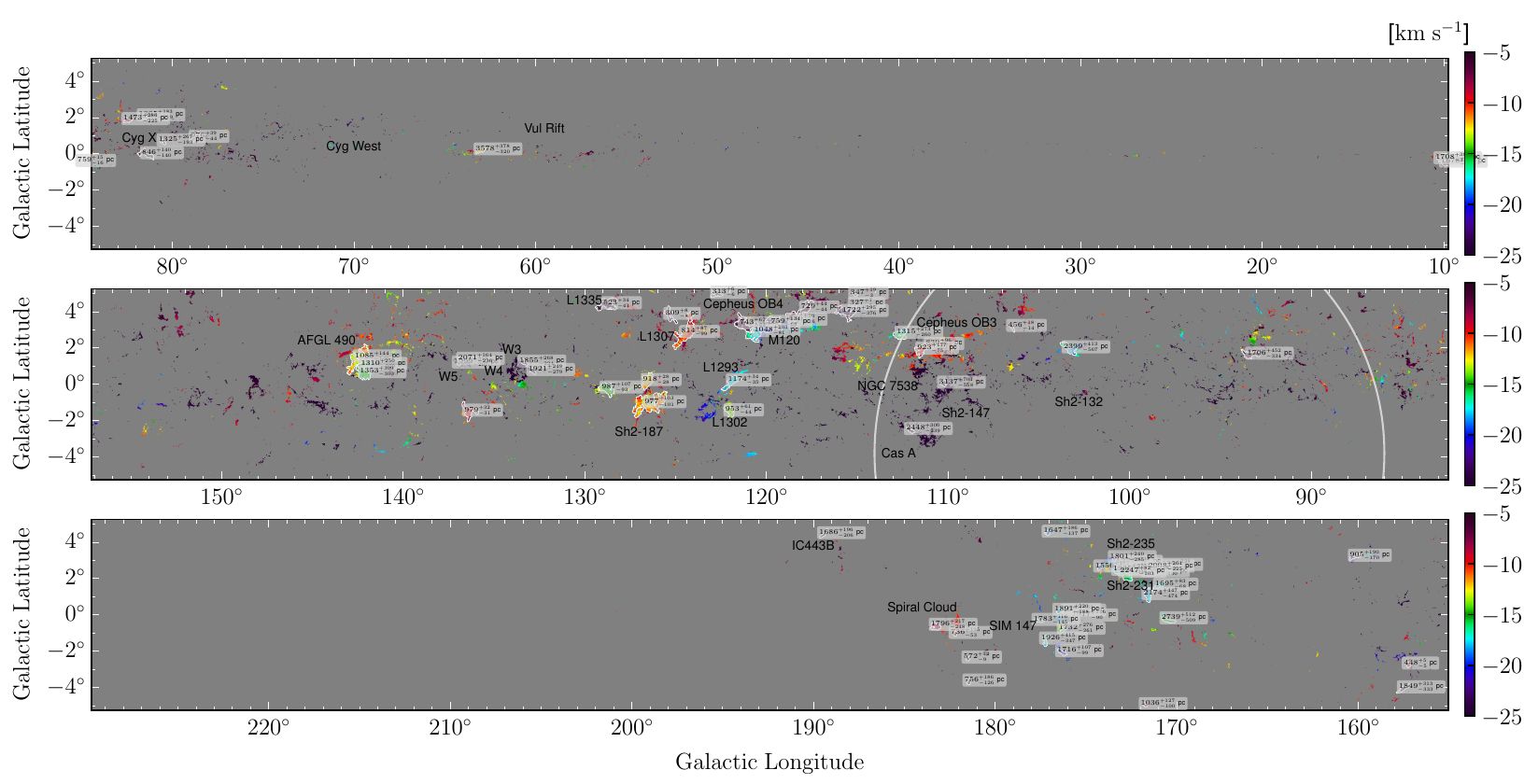}
\caption{Similar to Figure~\ref{fig:fig6}, but showing $^{13}$CO cloud structures with centroid velocities $v_{\rm LSR} < -5$ km s$^{-1}$. To better display the results, less negative velocity components are plotted in the top, and the color bar is limited to the range of $(-25, -5)$ km s$^{-1}$. The velocity components with more negative values are displayed in dark blue. }
\label{fig:fig8}
\end{figure*}

Nearby low-velocity MCs in the inner Galaxy extend to $l \sim 60^{\circ}$. While \citet{2023ApJS..267...30L} and \citet{2025AJ....170..239Z} have conducted a comprehensive analysis of the Vulpecula region, it lacks accurate distance measurements based on dust extinction methods, particularly in the solar neighborhood. The radial velocities of near clouds with distances ranging from $D \sim 400$ to $800$ pc in this region are generally high, with some exceeding $15$ km s$^{-1}$ (see Figure~\ref{fig:fig7}). As discussed later, these structures, together with the River MC, form another bubble-like structure with a diameter of approximately $200$ pc in the 3D distribution (see Figure \ref{fig:fig9} and Section \ref{split}).

MCs toward the Cygnus~West region suffer from severe velocity confusion near the tangent point, where the nearest clouds are concentrated at a distance of $\sim 1$ kpc. Together with the Cygnus Rift and Cepheus Spur (Section \ref{v3}), they comprise a $\sim 1$ kpc molecular gas overdensity oriented perpendicular to the LOS, although with discontinuities at $l \sim 72^{\circ}$ and $l \sim 100^{\circ}$ (see Figure \ref{fig:fig10} and Section \ref{cyg}). The molecular gas toward Cygnus has been systematically investigated in \citet{2024AJ....167..220Z}. Here, we find the Cygnus Rift extends eastward all the way to the direction of Cygnus OB7. And for the first time, we successfully measured the extinction-based distance of DR21 and W75N at $1506^{+306}_{-252}$ pc and $1231^{+125}_{-124}$ pc, respectively. In terms of mass, the $^{13}$CO clouds exhibit peaks at distances of $800\text{--}1000$ pc, $1.2$ kpc, $1.5$ kpc, and $1.7$ kpc, revealing a multilayer structure along the Local Arm. This is consistent with the detection of multiple jump points. 

Toward the Cyg OB7 region, there exists a large angular size MC structure in front of Cygnus Rift with its main body at $D \sim 600\text{--}700$ pc, and discrete clouds are distributed in the foreground at $D \sim 300\text{--}600$ pc. Together with foreground MCs which are predominantly distributed from within 200 pc to $\sim$500 pc and are spatially uniform at positive Galactic latitudes toward Cepheus, Cassiopeia, and Camelopardalis, these nearby structures are likely associated with the Gould Belt and the Radcliffe Wave \citep{2020Natur.578..237A}. Toward the Mon OB1 region, the distances of the two main clouds show differences: the eastern cloud is slightly farther than the western cloud. This result is consistent with \citet{2024ApJ...966..202Z}, although the distance measurements are systematically larger here.

\subsubsection{The high-positive-velocity MCs}\label{v2}

Most MCs in the inner Galaxy fall within this velocity interval (see Figures~\ref{fig:fig3} and \ref{fig:fig7}), exhibiting characteristics of high velocity and large distance. Similarly, to ensure clarity and uniformity, we present part of the clouds’ distances in Figure \ref{fig:fig7}. In the inner Galaxy, while a large number of MCs are concentrated in crowded, narrow ranges of velocity and latitude, distance measurements have been successfully obtained for most higher-latitude ($|b| > 1^\circ$), lower-velocity clouds. These clouds appear as fountain-like structures emerging from the Galactic disk, reaching vertical heights in excess of 100~pc and thus roughly beyond the typical thin disk scale height \citep{2021ApJ...910..131S}.

Nevertheless, we successfully measured distances to MCs associated with star-forming regions such as Serpens OB2, M17, M16, NGC 6604, and W44. Some structures appear at even greater distances, potentially located in the Norma Arm within $l \lesssim 30^{\circ}$. In the range $l \sim 56^{\circ}\text{--}62^{\circ}$, we detected a series of distant MCs extending continuously from $\sim 1.5$ kpc to over 4 kpc along the LOS. This is consistent with the location of the Local Spur traced by masers in \citet{2016SciA....2E0878X}, with velocities concentrated between 30 and 40 km s$^{-1}$.

The second quadrant yields almost no distance measurements listed in Table~\ref{table:tab1}. In the third quadrant, the Rosette and Maddalena clouds have been studied extensively in the context of star formation \citep{2010ApJ...724..687L, 2018ApJS..238...10L, 2024ApJ...971...14S}. The $^{13}$CO emission from Maddalena is very weak, resulting in insignificant detections here. Finally, MCs toward CMa OB1 exhibit a range of distances from 1.1 kpc to 1.7 kpc in this work.

\subsubsection{The high-negative-velocity MCs}\label{v3}

High-negative-velocity clouds are predominantly concentrated in the second quadrant. The most prominent concentration is found at $l \sim 110^{\circ}\text{--}150^{\circ}$, spanning Cepheus, Cassiopeia, the W3/4/5 complex, and extending to the AFGL 490 region. This corresponds to the Cepheus Spur structure at a distance of approximately 1 kpc (see Section \ref{bird}). The Cepheus Spur structure has been revealed in both OB star distributions and extinction maps \citep{2021MNRAS.504.2968P, 2025AJ....170..185Z}. Our results demonstrate that it is also clearly traced by $^{13}$CO molecular gas (see the global structure in the next section). In the western part of the region ($l \sim 100^\circ\text{--}115^\circ$), \citet{2021ApJS..254....3M} conducted a comprehensive study to MCs using MWISP $^{12}$CO and $^{13}$CO, and \citet{2026ApJS..284...42L} investigated the coevolution of a ``snake"-like stellar structure and its associated MCs. 

Among these MCs, distances to Sh2-187 \citep{2024AcASn..65...45X} and M120 \citep{2023RAA....23a5019S} have been precisely measured previously, while several other structures are spatially associated with Cepheus OB3 and OB4 at consistent distances. More distant cloud structures reside on the Perseus Arm, such as the MCs toward the W3/4/5 \citep{1998ApJ...502..265H, 2019ApJS..242...19L} and Sh2-235 \citep{2024MNRAS.527.4297C} regions, which have been comprehensively analyzed with CO and multi-band observations (e.g., sub-milimeter and near-far infrared data). Additionally, MCs on both sides of SIM 147 toward the anti-Galactic-center direction exhibit strong $^{13}$CO emission. Due to kinematic distance ambiguities in this region, some clouds in the third Galactic quadrant display large negative LOS velocities. Located at $1.5$--$2$ kpc within $180^\circ < l < 190^\circ$, these clouds encompass structures associated with IC 443 \citep{2003A&A...408..545W, 2014ApJ...788..122S} and a newly discovered ``Spiral cloud" that exhibits spiral-arm morphology with kinematic oscillations \citep{2026AJ....172....4C}. Together with additional low-velocity clouds (e.g., toward the Gem OB1 and GGD 4) in the outer Galaxy, these anti-Galactic-center structures appear as dense cloud complexes and constitute a significant portion of a coherent giant structure within the Perseus Arm (see Section \ref{bird}).

\subsection{Bird's-eye View of MCs' 3D distribution in Solar Neighbourhood}
\label{bird}

\begin{figure*}[ht!]
\includegraphics[width=18cm,angle=0]{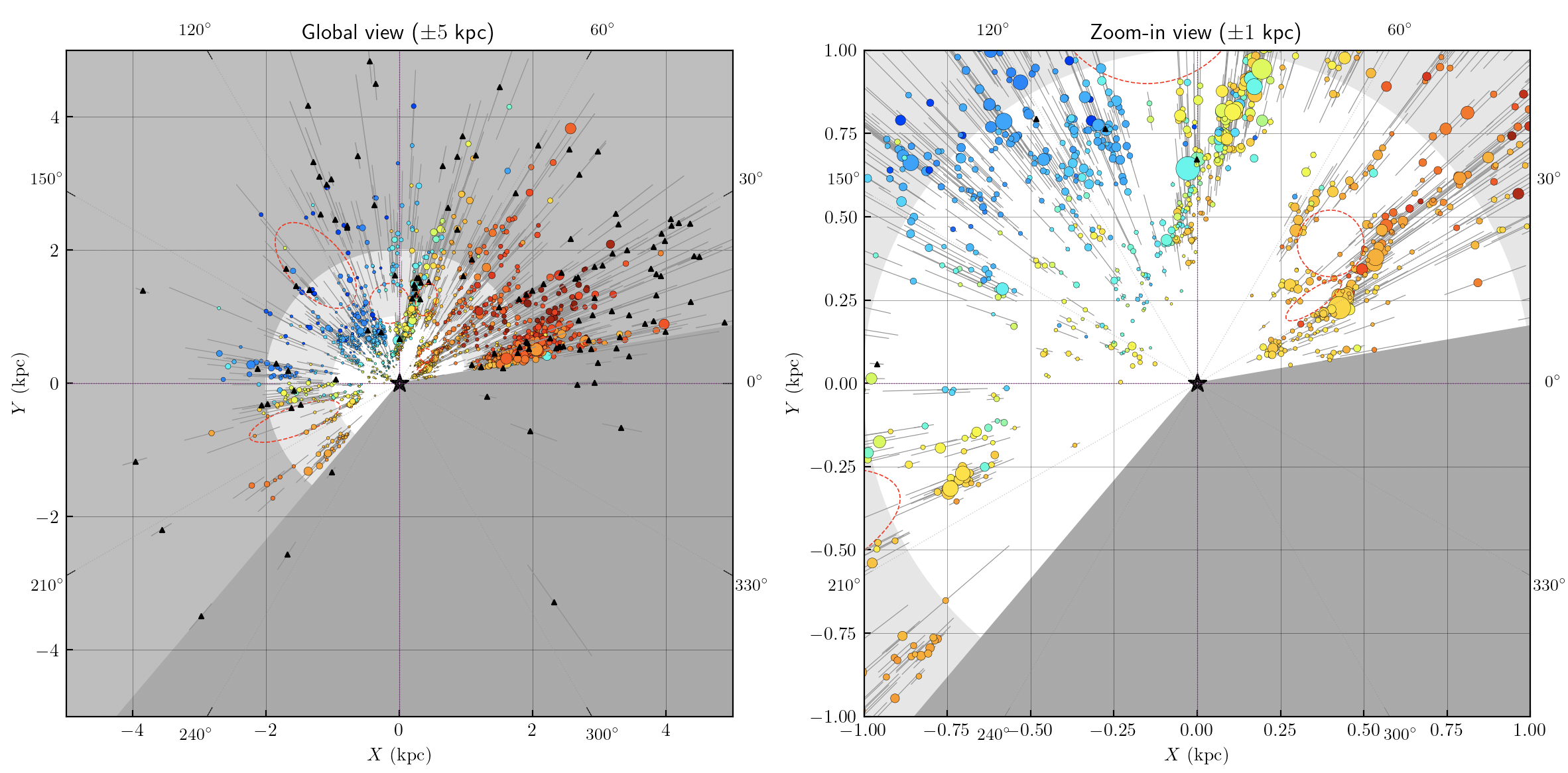}
\caption{Distribution of the 1751 MCs with measured distances in the solar neighborhood of the northern sky. 
\textit{Left panel:} Global view. \textit{Right panel:} Zoom-in view within 1 kpc. 
The circles represent the MCs, color-coded by their radial velocities according to the same color bar used in Figure~\ref{fig:fig3}. To improve visibility, the size of each circle is proportional to the square root of the cloud mass. The error bars denote the 16th and 84th percentiles of the bootstrap distance distribution. Maser sources from \citet{2019ApJ...885..131R} are similarly plotted as black triangles with error bars. The white background highlights the region within 1 kpc. The 1--2 kpc range is shaded light gray, while regions beyond 2 kpc are shown in dark gray. Areas not yet covered by the MWISP survey footprint are indicated by the deepest gray shading. The Sun is marked by a black five-pointed star at (0, 0). The superbubbles/cavities identified in related works (see Table \ref{tab:structures}) are highlighted by red dashed ellipses.}
\label{fig:fig9}
\end{figure*}

We project the cloud structures listed in Table~\ref{table:tab1} onto the $X\text{-}Y$ Cartesian coordinate system. Left panel of Figure~\ref{fig:fig9} presents the global distribution within 5 kpc, while the right panel zooms into the solar neighborhood within 1 kpc. Following the quantitative assessment of distance uncertainties in Section \ref{err}, we group the cloud sample at different heliocentric distances. As shown in Figure \ref{fig:fig9}, we adopt a three-tiered grayscale scheme: white for the ``gold" 1 kpc sample characterized by small errors and high completeness, light gray for the 1--2 kpc regime, where uncertainties increase but the sample remains robust, and dark gray for distances beyond 2 kpc, where absolute distance errors are large and incompleteness becomes severe. Regions outside the MWISP coverage are masked with the darkest shade. This partitioning provides a critical framework for evaluating the reliability of large-scale structures identified later. Specifically, the distance error bars associated with each cloud serve as a visual proxy for the robustness of its assigned position. 

The distribution of $^{13}$CO-traced MC structures does not strictly follow the pattern of the Galactic spiral arms in Solar neighbourhood. Instead, they exhibit numerous linear structures within the solar neighborhood. We interpret these CO cloud features as reflecting a remarkably inhomogeneous molecular gas distribution on large scales. Giant bubble-like structures are the most prominent features, intertwining with the linear structures and contributing to the complexity of the molecular gas distribution in the solar neighborhood. Here we present the $^{13}$CO-traced kpc-scale linear features, cavities, and newly identified sub-kpc structures, ordered by increasing Galactic longitude.

\begin{figure*}[ht!]
\plotone{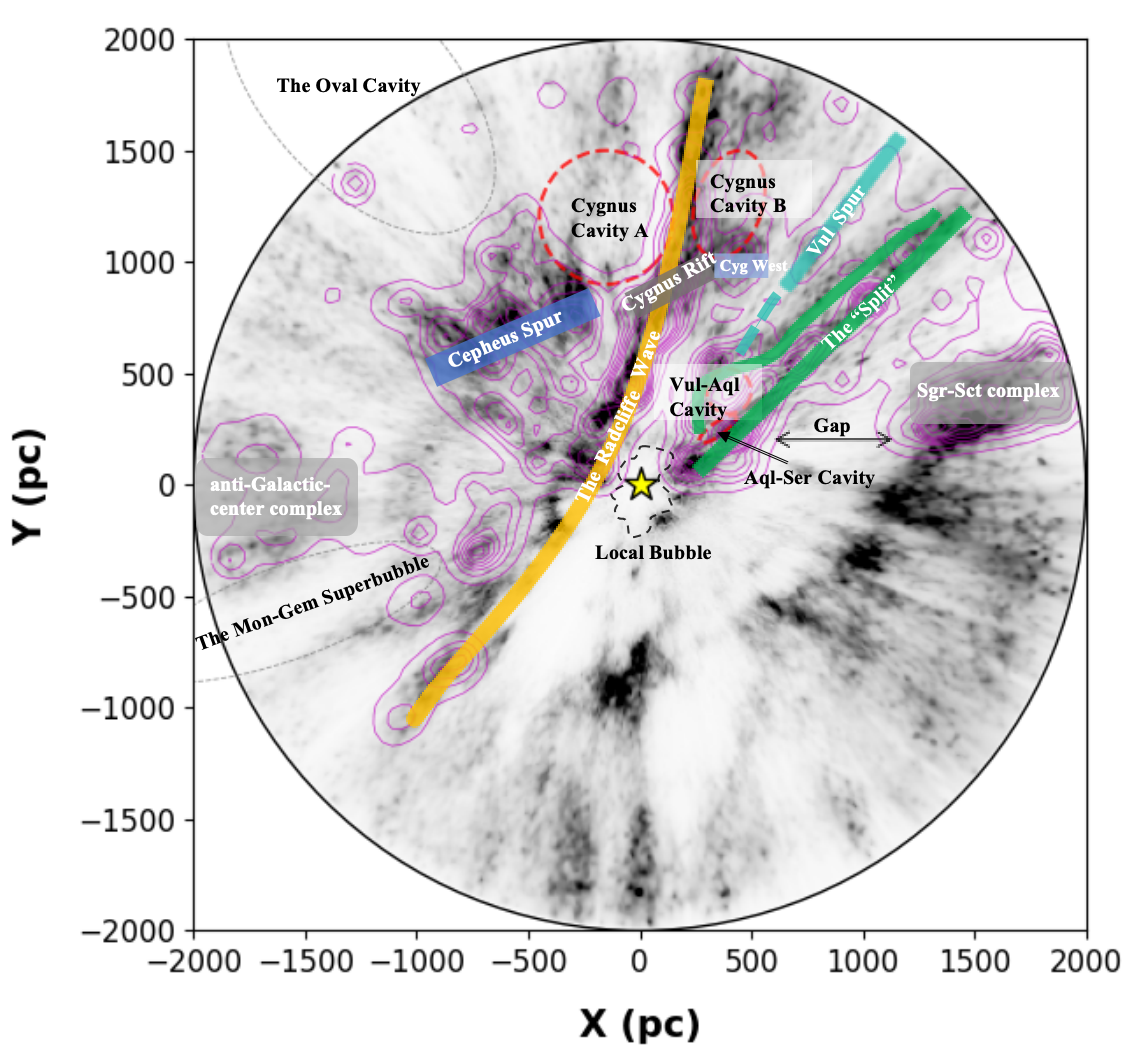}
\caption{Roadmap of the major large-scale structures within 2 kpc of the solar neighborhood. The background grayscale image shows a 3D dust extinction reconstruction in the $X$--$Y$ plane, derived from Gaia XP spectrophotometric reddening \citep{2023MNRAS.524.1855Z} using the CU-JADE multi-jump detection algorithm. The integration range is $|Z| < 300$ pc. The magenta contours represent the equivalent density map of $^{13}$CO clouds from the MWISP survey (Table \ref{table:tab1}) in the northern Galactic plane. We overlay the kpc- and sub-kpc-scale structures discussed in Section \ref{sect5}. These structures correlate spatially with both the dust and $^{13}$CO-traced gas. Closed dashed curves (ellipses, circles, and irregular shapes) denote large-scale cavities or bubbles, with newly identified cavities highlighted in red. Colored bands indicate large-scale linear or spur-like features. The names of these structures are labeled at their respective positions. }
\label{fig:fig10}
\end{figure*}

The first notable feature is the gap observed in the $l\text{-}d$ (Figure~\ref{fig:fig3}) and $X\text{-}Y$ (Figures~\ref{fig:fig9} and \ref{fig:fig10}) projections for $l < 25^{\circ}$. This gap appears to represent the inter-arm region between the Aquila-Serpens complexes at Local Arm and the Sagittarius Arm (Sgr), extending from approximately 500 pc to 1.2 kpc (named Aql--Sgr Gap in Table \ref{tab:structures}).

However, the Sagittarius and Scutum-Centaurus (Sct) arms do not exhibit clear separation (Figure \ref{fig:fig9}). Within $l < 45^{\circ}$, cloud mass is particularly concentrated in the distance range of $1.5\text{--}3$ kpc, where clouds accounting for 56.1\% of the sample contain approximately 82.5\% of the total mass. In the face-on view, this manifests as a dense concentration of MCs elongated along the LOS (named Sgr--Sct Complexes). This phenomenon is possibly attributable to three factors.
First, the Sgr and Sct arms are physically close in the low-longitude range. Studies by \citet{2025AJ....170..185Z} and \citet{2021A&A...653A..33N} place the Sgr and Sct arms at $\sim 1.6$ kpc and $2.8$ kpc, respectively. Considering the width of the spiral arms, this $\sim 1$ kpc separation is small. Second, distance measurement uncertainties play a significant role. We statistically analyzed the distance errors for clouds in this region ($l < 45^{\circ}$, $D$ in $[1.5, 3]$ kpc) using the CU-JADE error propagation method. The mean relative error ($\delta d / d$) is 0.163, corresponding to a mean distance uncertainty of 335 pc, which is sufficient to blur the distinction between arm and inter-arm gas. Finally, this could also be a manifestation of gas pile-up driven by certain Galactic dynamical mechanisms, which renders the spiral arm structure in this region rather indistinct. This direction points toward the near side of the Galactic bar, where bar-driven perturbations combined with disk rotation facilitate the formation of such dense concentrations. Indeed, high-precision maser parallax measurements show a similar degree of confusion (Figure~\ref{fig:fig9}). 

In the right panel of Figure~\ref{fig:fig9}, the foreground Aquila-Serpens complexes appear as a tunnel-like structure in the face-on view, connecting the low-density region of the Local Bubble on the near end and showing an accelerating velocity trend toward the Phoenix MC on the far end. Adjacent to it to the east is a larger, roughly circular cavity located at $\sim 570$ pc with a diameter of $\sim 200$ pc. Both cavities lie on the near side of the gaseous ``Split” (see Section \ref{split} for details). Together with the Split's substructures, the radially extended linear feature toward Vulpecula Rift (named Vul Spur here; Figure \ref{fig:fig10}), and the Local Spur, they delineate a rich network of large-scale linear structures in the inner Galaxy ($l \lesssim 60^\circ$) near the Sun, with typical lengths exceeding 1 kpc.

The most prominent linear gas structures extending over $2$ kpc from the Cygnus~X region (and possibly farther) through NGC 7000 and Cyg OB7, reaches to the near side to the Sun (see Figures \ref{fig:fig9} and \ref{fig:fig10}). Note that our data do not fully trace the component associated with the Gould Belt at higher Galactic latitudes, the molecular gas traced by $^{13}$CO does not extend to CMa OB1 region. Figure~\ref{fig:fig10} presents the CU-JADE–based 3D dust reconstruction, in which the linear structure aligns with the Radcliffe wave \citep{2020Natur.578..237A}. 
The Radcliffe wave is a $\sim 2.7$ kpc structure tracing the Local Arm, often regarded as its gas reservoir \citep{2022A&A...664L..13S}. Some studies suggest it exhibits wave-like characteristics in 3D distribution and kinematics 
\citep{2022A&A...660L..12T, 2024ApJ...971..167Z}, forming part of larger Galactic wave patterns. Our results confirm that the Radcliffe wave contains a substantial fraction of the molecular gas in the solar neighborhood. Due to space limitations, detailed discussion of the Radcliffe Wave and the Local Spur is deferred to future work. 

Toward the Cygnus direction of the Radcliffe wave, we discover two large sub-kpc-scale cavities flanking the wave. We will elaborate on their relationship with the Cygnus Rift, the Cepheus Spur, and the foreground MCs in Cygnus~West in Section~\ref{cyg}.

In the region defined by $110^{\circ}<l<150^{\circ}$ and $0.65<d<1.25\,\mathrm{kpc}$,
a top-down view reveals an additional MC over-density. The majority of these clouds exhibit negative LOS velocities, distinct from these foreground MCs. Based on its spatial coincidence with the CU-JADE 3D dust reconstruction and its consistency with \citet{2025AJ....170..185Z}, we attribute the $^{13}$CO-traced gas extension to the Cepheus Spur \citep{2021MNRAS.504.2968P}. 

Toward $l \sim 130^{\circ}$, the $^{13}$CO clouds outline a cavity consistent with the ellipse feature provided by \citet{2025AJ....170..185Z}, which is delineated based on CU-JADE extinction maps (overlay in Figures~\ref{fig:fig9}, \ref{fig:fig10} and \ref{fig:figb2}). Specifically, the MCs at $l \sim 110^{\circ}$ beyond 2 kpc, the W3/4/5-associated complexes, and the far-side Cepheus Spur clouds lie on the periphery of this cavity. The giant bubble structure (Oval Cavity) was first identified in the extinction maps of \citet{2022A&A...664A.174V}. The originally proposed scale was larger than \citet{2025AJ....170..185Z}, encompassing maser-traced HMSFRs and $^{13}$CO/$^{12}$CO clouds within its interior (see Figure \ref{fig:figb2}). The presence of this bubble disrupts the continuity of the Perseus Arm in the second quadrant. Its age, formation, and kinematic properties are investigated in \citet{2025NatCo..1610558C} using OB star samples.   

In the anti-Galactic-center region, as seen in Figure \ref{fig:fig9}, the MCs in the Perseus Arm exhibit radial velocities from negative to positive across the second to third quadrants, which is attributed to differential Galactic rotation, forming a molecular complex (Figure \ref{fig:fig10}). Distances to MCs toward the anti-Galactic-center have long suffered from substantial uncertainties. Here, we present the first comprehensive sky mapping and top‑down perspective, aimed at resolving their connection to the Perseus Arm, located at $\sim 2$ kpc but exhibiting a LOS velocity spread exceeding 30 km s$^{-1}$. Adjacent to this complex lies another kpc-scale bubble. 

In the third quadrant, we identified and named the Mon-Gem superbubble in \citet{2025AJ....170..185Z}, which is also prominently traced by $^{13}$CO clouds (see Figure \ref{fig:fig9}). This reveals a vast cavity approximately $1.5$ kpc in extent, located beyond the Mon OB1 region. No central energy source sufficient to drive such a large structure is found. Therefore, similar to the giant bubble at $l \sim 130^{\circ}$, it is likely a very old cavity that has undergone slow expansion over an extended period. 

Some of these large-scale structures
may be artifacts caused by heliocentric-distance uncertainties and LOS elongation (i.e., finger-of-God-like radial elongation effect, particularly in the inner Galaxy), while others, such as the Radcliffe wave \citep{2020Natur.578..237A} and the Local Spur, are reexamined based on our CO study (Figure \ref{fig:fig9}). Additionally, several large-scale structures remain insufficiently studied, including the ``Split" structure \citep{2019A&A...625A.135L} and the gaseous Cepheus Spur extension \citep{2021MNRAS.504.2968P}. 

All large-scale molecular gas structures traced by $^{13}$CO in this work are cataloged in Table \ref{tab:structures}, several of which are reported here for the first time. We note that this work is anchored in $^{13}$CO-traced cloud structures. Among the structures mentioned in this paper and Table~\ref{tab:structures}, apart from the 3D dust itself, all others were first identified from the $^{13}$CO cloud catalog with precise distances, with the 3D dust extinction map serving only as a check and supplement. Upon inspection, every molecular gas structure has a counterpart in the 3D dust extinction map. Additionally, we mention via private communication with Wang et al.\ (accepted in \textit{Nature}) that their independent 3D extinction map produces structures consistent with ours.
In the following section, we delve into the newly identified structures at finer spatial scales and discuss their properties.

\begin{table}[htbp]
\centering
\caption{Gaseous and dust structures within 5\,kpc of the Sun.}
\label{tab:structures}
\begin{tabular}{llll}
\toprule
Object Type & Name & Information & Reference \\
\midrule
Molecular Cloud &
MWISP Phase I $^{13}$CO cloud structures &
1751 clouds in Table \ref{table:tab1} &
This work \\
\midrule
\multirow{2}{*}{Overdensities} &
Sgr--Sct complexes &
$10^{\circ} < l < 25^{\circ},\; 1.5 < d < 3\,\mathrm{kpc}$ &
This work \\
& Anti-Galactic-center complexes &
$170^{\circ} < l < 190^{\circ},\; 1.5 < d < 2.5\,\mathrm{kpc}$ &
This work \\
\midrule
Dust &
3D dust map &
Resolution $10\,\mathrm{pc} \times 10\,\mathrm{pc}$ within 2\,kpc &
\textsuperscript{a} \\
\midrule
\multirow{6}{*}{Spur-like Structures} &
$^{13}$CO-traced Radcliffe Wave &
$(x,y)=(-1.38,-1.35)$ to $(0.96,2.8)\,\mathrm{kpc}$ &
This work; \textsuperscript{b} \\
& $^{13}$CO-traced Split &
$(x,y)=(0.2,0.1)$ to $(2.4,1.6)\,\mathrm{kpc}$ &
This work; \textsuperscript{c} \\
& $^{13}$CO-traced Local Spur &
$(x,y)=(0.9,1.65)$ to $(2.6,3.8)\,\mathrm{kpc}$ &
This work; \textsuperscript{d} \\
& $^{13}$CO-traced Cepheus Spur &
$110^{\circ}<l<150^{\circ},\; 0.65<d<1.25\,\mathrm{kpc}$ &
This work; \textsuperscript{e} \\
& $^{13}$CO-traced Cygnus Rift &
$72^{\circ}<l<95^{\circ},\; 0.7<d<1.1\,\mathrm{kpc}$ &
This work; \textsuperscript{a, f} \\
& Vul Spur &
$(x,y)=(0.7,0.9)$ to $(2.0,2.4)\,\mathrm{kpc}$ &
This work \\
\midrule
\multirow{5}{*}{Superbubbles / Cavities} &
Oval Cavity &
$(\ell,b,d,D)=(125^{\circ},0^{\circ},2.17\,\mathrm{kpc},22^{\circ})$ &
This work; \textsuperscript{g, a} \\
& Mon--Gem Superbubble &
$(\ell,b,d,D)=(200^{\circ},-3^{\circ},1.67\,\mathrm{kpc},14^{\circ})$ &
This work; \textsuperscript{a} \\
& Cygnus Cavity A &
$(\ell,b,d,D)=(97^{\circ},-4^{\circ},1.21\,\mathrm{kpc},28^{\circ})$ &
This work \\
& Cygnus Cavity B &
$(\ell,b,d,D)=(72^{\circ},0^{\circ},1.32\,\mathrm{kpc},13^{\circ})$ &
This work \\
& Vul--Aql Cavity &
$(\ell,b,d,D)=(46.5^{\circ},0^{\circ},580\,\mathrm{pc},19.5^{\circ})$ &
This work \\
& Aql--Ser Cavity &
$(\ell,b,d,D)=(35.5^{\circ},0^{\circ},430\,\mathrm{pc},10^{\circ})$ &
This work \\
\midrule
Others &
Aql--Sgr Gap &
$10^{\circ}<l<25^{\circ},\; 0.5<d<1.2\,\mathrm{kpc}$ &
This work \\
\bottomrule
\end{tabular}

\medskip
\raggedright
\footnotesize
\textit{Note.}
--- All structures proposed in this paper are primarily supported by the $^{13}$CO cloud structure catalog. For the large-scale structures newly discovered in the $^{13}$CO catalog, especially those within 2~kpc, we use the extinction map (integrating dust over $|Z| < 300$\,pc) produced by CU-JADE as a complementary tracer to check potential sample biases (e.g., Cyg Cavity~A; Section~\ref{cyg}), given that the CO footprint covers $|b| \leq 5^{\circ}.15$. \\
---Cartesian coordinates $(x,y)$ are given in the Galactic plane, with the Sun at the origin. $(\ell,b,d,D)$ denote longitude, latitude, heliocentric distance, and angular diameter, respectively. \\
$^{a}$ \citet{2025AJ....170..185Z}; $^{b}$ \cite{2020Natur.578..237A}; $^{c}$ \citet{2019A&A...625A.135L}; $^{d}$ \citet{2016SciA....2E0878X}; $^{e}$ \citet{2021MNRAS.504.2968P}; $^{f}$ \citet{2024AJ....167..220Z}; $^{g}$ \citet{2022A&A...664A.174V}. 
\end{table}

\subsection{Interesting sub-regions in four-dimensional map}
\label{subregion}

\subsubsection{Cavities toward the Split} \label{split}

In this section, we dissect the substructure of the gaseous Split. First, the Split is not simply a split. Though it appears as a prominent inter-arm structure in dust extinction maps, $^{13}$CO-traced molecular gas unveils a far more complex structure. Within $\sim 1$ kpc of the Sun, where distance measurements are precise and the completeness is high, the fine structure of the local molecular gas is clearly resolved, corresponding to the foreground Serpens Near and Main, River, and Vulpecula Rift MCs
described in Section~\ref{v1}.

We identify two prominent neighboring cavities that shape the near side of the Split. One is located at $l \sim 35^{\circ}$ (named Aql--Ser Cavity) at a distance of $\sim 450$ pc, which is associated with the Serpens, River, and Phoenix MCs mentioned earlier. The other adjacent cavity is located at $l \sim 47^{\circ}$ (named Vul--Aql Cavity) and lies at a distance of $\sim 600$ pc. 

\begin{figure*}[ht!]
\plotone{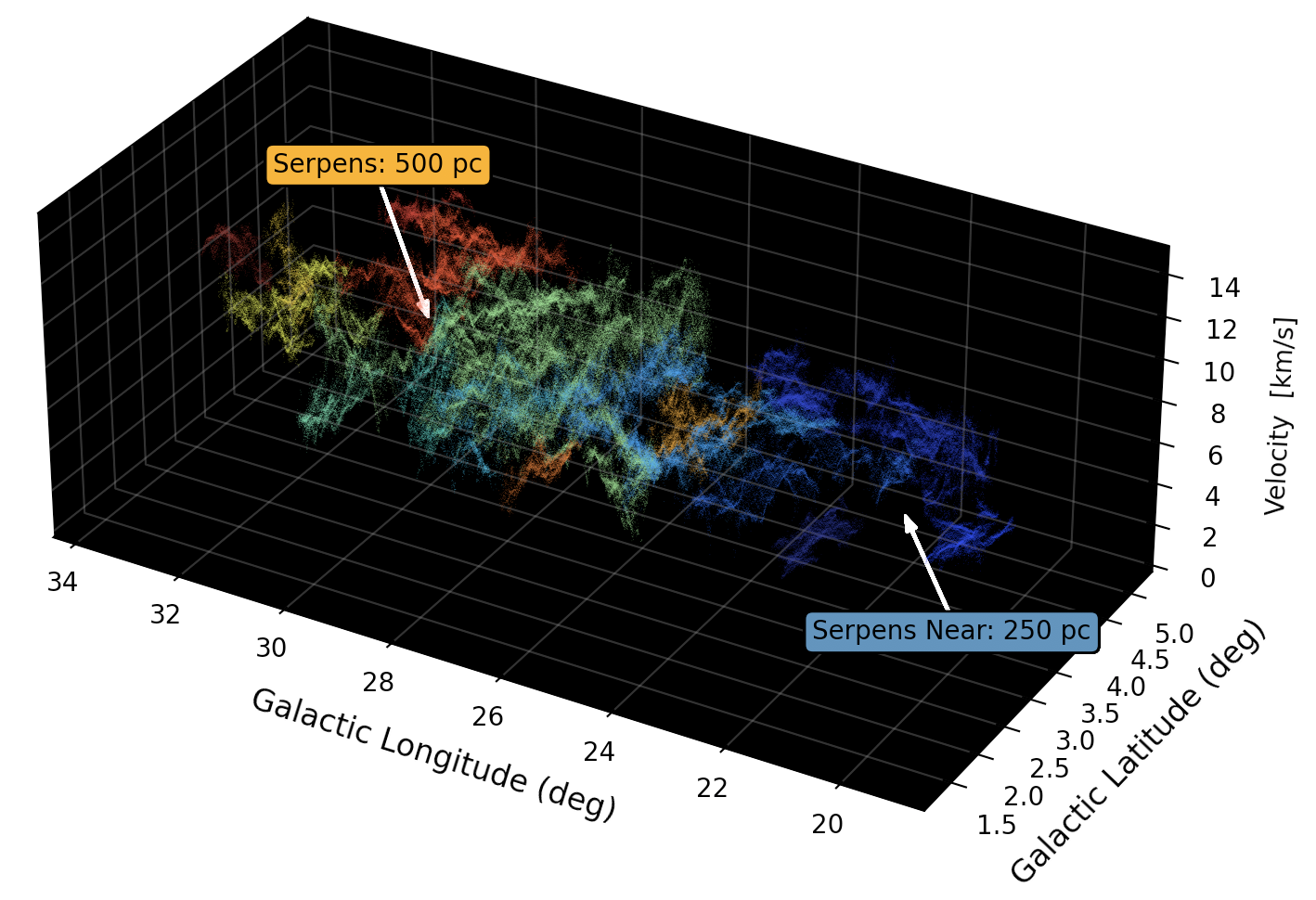}
\caption{Pixel-by-pixel scatter plot of $^{13}$CO MCs toward the Serpens Near ($236\text{--}261$ pc; blue) and Serpens Main ($436\text{--}550$ pc; green to red) in PPV space. The clouds overlap spatially at Galactic longitudes $l \sim 26^{\circ}\text{--}28^{\circ}$, while their bulk velocities are concentrated in the range of $4\text{--}10$ km s$^{-1}$. Different colors indicate the distance distribution of the clouds, ranging from $236$ pc to $550$ pc.}
\label{fig:fig11}
\end{figure*}

The Aql--Ser Cavity is spatially contiguous with the Local Bubble. We note that the Aql--Ser Cavity extends from its near side,  which coincides with the edge of the Local Bubble at a heliocentric distance of $\sim 200$ pc, to its far-side at the Phoenix MC at $\sim 650$ pc.  Consequently, its kinematic signature (i.e., velocity gradient toward the Phoenix cloud) and its structural properties (i.e., extension into an elongated cavity) are consistent with a connection to the Local Bubble and its energy sources, specifically the Lower Centaurus Crux (LCC) and Upper Centaurus Lupus (UCL) in Scorpius-Centaurus OB association \citep{2022Natur.601..334Z}. 
After accounting for the $\sim 6~\mathrm{km~s^{-1}}$ LOS velocity shift due to Galactic differential rotation \citep[using constant $A = 15~\mathrm{km~s^{-1}~kpc^{-1}}$;][]{1986MNRAS.221.1023K}, the residual radial velocity across this cavity is $\sim 4~\mathrm{km~s^{-1}}$. Assuming a continuous energy injection from the Local Bubble via shocks and stellar winds, and that the Phoenix MC has been displaced by $\sim 200$ pc at a time-averaged travel speed (the residual $v_{\rm LSR}$ mentioned above), the implied timescale is approximately several tens of Myr. This result is within an order-of-magnitude of the age of the Local Bubble, which aligns reasonably well with models of sustained supernova explosions over the past 14 Myr \citep{2022Natur.601..334Z}.

Figure~\ref{fig:fig11} provides further evidence for this scenario. It shows that from $l \sim 20^{\circ}$ toward the center of the cavity, the centroid velocities of the clouds exhibit increasingly intense oscillations with increasing distances. These oscillations are thought to arise from the combined effects of self-gravity and large-scale external turbulent flows \citep{2020NatAs...4.1064H}. Interestingly, recent work \citep{2026arXiv260605668R} has identified 88 new Herbig-Haro (HH) objects in the vicinity of the Aquila-Serpens complex. These HH objects originate from spatially distinct, actively star-forming clouds, consistent with the the compressed cloud structures along the edge of the Aql--Ser Cavity identified in this work. The spatial association suggests that star formation near the bubble boundary may be triggered by bubble expansion \citep{1998ApJ...498..689H, 2022Natur.601..334Z, 2026AJ....172...98S}. 
Alternatively, the Aql--Ser Cavity resembles a chimney extending along the Galactic plane, analogous to the northern Local Chimney \citep{2003A&A...411..447L}, which is rather expanded vertically toward Galactic halo. Simulations suggest that such persistent supernova activity can drive cavities to break out of the Galactic disk \citep{2022ApJ...924L..28O}. In this scenario, the northern chimney, potentially powered by the sequential explosion of $\sim 15$ massive stars over $\sim 14$ Myr, extends vertically from z $\sim 150$ pc to z $\sim 600$ pc \citep{2022Natur.601..334Z, 2024ApJ...973..136O}.

While the Split lies in the background of the Aql--Ser Cavity, with its far side extending toward and gradually merging with the Sagittarius Arm, another distinct branch toward the Vulpecula Rift (i.e., the Vul Spur) acts as a spur-like structure running parallel to the Radcliffe Wave and the Local Spur. The presence of the Vul--Aql Cavity imparts a distinct distortion to the $^{13}$CO clouds on both sides. At a distance of approximately 1.3 kpc in the direction, a series of MCs bridges these two linear structures. This cavity feature is also discernible in the extinction map of \citet{2022A&A...664A.174V}, although its formation mechanism remains unclear.

\subsubsection{Cavities toward Cygnus Region}
\label{cyg}
We identify a nearly circular bubble (named Cygnus Cavity A) situated between the Cepheus Spur and Cygnus~X, with a diameter of approximately 600 pc and a distance of about 1.2 kpc from the Sun (Figures \ref{fig:fig9} and \ref{fig:fig10}). Although smaller than the aforementioned kpc-scale bubbles, it is still significantly larger than any known giant gas bubbles in the solar neighborhood driven by stellar winds or supernova remnants. We further employ the 3D extinction map generated by CU-JADE \citep{2025AJ....170..185Z} to trace this cavity at varying vertical heights $z$. We find that even at this scale, it does not break out of the Galactic dust disk, and its center is located $\sim 80$ pc below the Galactic plane (see Figures \ref{fig:fig6} and \ref{fig:fig10}). A detailed characterization of its 3D structure is beyond the scope of this paper. Intriguingly, the presence of this bubble appears to coincide with a discontinuity in the Cepheus Spur and Cygnus Rift near $l \sim 100^{\circ}$. Several maser sources are located in its vicinity (see Discussion).

Another bubble (Cygnus Cavity B) is identified on the opposite side of Cygnus~X at $l \sim 72^{\circ}$. Its near and far end appear to be open, giving it the appearance of a strait rather than a closed bubble, with a major axis of $\sim 500$ pc and a minor axis of $\sim 300$ pc. Spatially, its presence appears to separate the Cygnus Rift from the foreground MCs in Cygnus-West. 

We cannot associate either of these bubbles with the Cygnus Superbubble \citep[the `C'-shaped structure observed in soft X-ray emission;][]{1980ApJ...238L..71C, 2020ApJ...905...91B}, as this high-energy source is centered on Cygnus~X itself rather than its flanking regions. It is generally believed to be driven by the stellar winds or sequential bursts from/within hundreds of OB stars in Cygnus OB2 and OB1 association. While these bubbles share similar distances, the Cygnus Superbubble leaves no discernible kinematic imprint on the face-on view of 3D dust extinction and MC distributions. Furthermore, the radial velocities of the MCs in the Cygnus region are too complex to provide conclusive evidence. Recently, we have identified spatial morphological evidence in $^{12}$CO and dust at higher Galactic latitudes (manuscript in preparation), which had not been thoroughly investigated previously due to sensitivity limitations and high extinction. Future studies may further study the relationship and potential interactions between these two bubbles and the Cygnus Superbubble.

\section{Discussion}
\label{discuss}

\subsection{Error Propagation and Sample Contamination}
\label{bs}
Propagating uncertainties from individual stellar measurements to a derived physical quantity (i.e., jump point) is non-trivial, particularly when complex mathematical relationships exist between the sample data (parallax-extinction) and the target parameter. Traditional single-jump methods often circumvent this by prioritizing extinction errors and adopting the median/mode of the distance distribution for each stellar sample. While fitting piecewise functions via MCMC, these methods typically omit the propagation of parallax uncertainties from individual stars. Consequently, a mean distance uncertainty of the stellar sample must be added post-process to the systematic error. Although reasonable, this approach implies that the systematic error varies with distance.

Given typical Gaia parallax uncertainties of $\sim 0.1$ mas \citep{2023A&A...674A...1G}, the relative distance uncertainty is approximately 10\% at 1 kpc, decreasing for nearer sources and increasing for farther ones. This prevents the assignment of a uniform uncertainty value. Furthermore, it highlights the rapid degradation of Gaia's distance precision with increasing distance. While precision is high ($\lesssim$10\%) within 1 kpc, yielding robust structures in face-on views, it deteriorates rapidly beyond this range. Researchers often impose a cutoff on relative parallax errors (e.g., $< 20\%$) for distant clouds, which further reduces the usable sample size. Combined with foreground extinction, this limits Gaia-based distance measurements of MCs to roughly 3 kpc.

The direct propagation of systematic errors on MCs also suggests that Gaia parallax measurements within a specific field are not entirely independent, and they share common uncertainty pattern. Statistics from Table~\ref{table:tab1} support these conclusions. This implies a pessimistic constraint: simply increasing the stellar sample size cannot effectively reduce the distance uncertainty at jump positions. Since stars within the same cloud structure are likely observed in the same field, repeated scanning reduces observational noise but cannot alter the fact that the measurements are not independent. The distance uncertainties from our CU-JADE method corroborates this. In principle, the distance uncertainty derived from CU-JADE reflects the uncertainty of the on-cloud stellar sample, scaling from $\sim 5\%$ at 500 pc to $\sim 20\%$ at 2 kpc (see Figure \ref{fig:fig45}). Beyond 2 kpc, the uncertainty ceases to increase monotonically and instead fluctuates, precisely because we filter the input data based on parallax uncertainty thresholds. 

Distance uncertainties that rival the physical size of a structure inevitably induce artificial LOS stretching in the face-on projection, effectively broadening the apparent depth of the structures. For clouds within 1 kpc, this distance-induced smearing is largely suppressed (right panels in Figures \ref{fig:fig9} and \ref{fig:figb2}), as the distance errors are minor compared to their projected extents. However, beyond 3 kpc, uncertainties typically surpass 500 pc, an order of magnitude larger than the typical giant MC scale. In this regime, the cloud distribution is stretched into spur-like filaments aligned with the LOS, rendering the identification of coherent structures highly ambiguous.

The advantage of the CU-JADE's error analysis is that it quantifies the precision of our results, allowing us to assess the reality of the large-scale structures composed of these MCs. Besides precision, there are other factors 
that affect distance accuracy, such as the reliability of cloud identification in complex regions and variations in dust/gas properties across different environments. Importantly, LOS confusion remains a challenge, especially for small-scale, clumpy clouds. 
For these samples, we impose additional constraints and conduct manual inspections (see Section \ref{catalog}).
Although we cannot guarantee the complete removal of all contaminated results (account for roughly 10\%), the consistency with maser parallaxes (Section \ref{accuracy}) and with other independent measurements that also rely on Gaia parallaxes (see Appendix \ref{appb}) demonstrates that our method is generally effective, particularly for resolving complex, crowded, and distant cloud structures along the Galactic plane via multi-jump point traversal. Given the recent release of Gaia data (DR4), we can also use more precise extinction-parallax measurements and an improved methodology for retrospective checks. 

\subsection{Completeness, Reality of Large-Scale Structures, and Intrinsic Connections}
\label{cri}
Following the discussion of uncertainties, we emphasize the high efficiency of our method. Even toward the inner Galaxy, distances are measured for MCs accounting for nearly 40\% of the total flux. Evidently, the bulk of this flux originates from nearby clouds with large angular scales. Although we cannot quantify the completeness as a function of distance, an examination of Figures~\ref{fig:fig5}, \ref{fig:fig45}, and the case analysis (e.g., Cygnus Rift) leads us to estimate that the census within 1 kpc is relatively high. The number recovery fraction drops significantly between 1 and 2 kpc, yet this does not appear to severely bias the uniformity of the large-scale structure distribution. However, beyond 2 kpc, the escalating distance uncertainties, compounded by the severe incompleteness of the distance-measured sample, dominate the analysis. Within this regime (the dark gray region in Figure \ref{fig:fig9}), the cloud catalog provides only limited constraints on the molecular gas distribution in the solar neighborhood.

Consequently, the Aql--Ser and Vul--Aql Cavities possess high reliability and exhibit no significant LOS distortion. The Aql--Ser Cavity is further supported by robust kinematic evidence, indicating that this region has experienced large-scale work done by shocks and stellar winds. The bubbles near Cygnus exhibit a complex, intertwined relationship with the spur-like structures. The two sub-kpc scale cavities show associations with the gaseous Radcliffe Wave, Cepheus Spur, Cygnus Rift, and foreground MCs toward Cygnus-West, with no apparent LOS elongation. 

In contrast, the two kpc-scale superbubbles show obvious distortion and incompleteness at their far sides. Nevertheless, their existence is corroborated by multiple independent tracers. The cavity toward $l \sim 130^{\circ}$ has been delineated by dust and expanding OB stars on Perseus Arm. The proximal segment of this structure, which lies adjacent to the Cepheus Spur, coincides with HMSFRs revealed by Cepheus OB3 and OB4 associations. Owing to the small heliocentric distance, the CU-JADE 3D extinction map resolves numerous small-scale bubbles within the Spur.
The Mon-Gem superbubble discovered toward $l \sim 200^{\circ}$ is also a prominent feature, warranting deeper investigation in future studies. The sightline toward Mon OB1 is dominated by its $\sim$800 pc cloud complex and lacks prominent background structures. Nevertheless, Mon OB1 complex itself harbors bubble-like cavities and coeval young stellar groups \citep{2022AJ....163..266L, 2024ApJ...964...93L}. 
Owing to minimal distance uncertainties and high detection completeness, the near sides of kpc-scale gas features yield the most reliable substructure censuses and stringent constraints on Galactic disk energy transfer, serving as critical laboratories for studying stellar feedback–ISM coupling from clouds to galactic structures.

We further highlight the indications of coexistence, dependence, influence, and even direct interaction among spurs, cavities, and molecular gas overdensities. This implies the profound and lasting impact of Galactic differential rotation and massive stellar feedback on the disk. We observe that superbubbles tend to cluster around spur or linear structures, while masers tracing HMSFRs tend to reside on the periphery of superbubbles (Figure~\ref{fig:fig9}). The picture is consistent with the scenario in which the nearby star formation appears at the boundaries of bubbles, as evidenced by the star formation at the Local Bubble's edge and the abundance of HH objects surrounding the Aql-Ser Cavity (see Section \ref{split}). This underscores the critical role of stellar and supernova feedback in shaping Galactic structure, sculpting density distributions, and potentially triggering the next generation of star formation. 

Leveraging the $\lesssim1$ kpc gold sample (right panel of Figures~\ref{fig:fig9} and \ref{fig:figb2}), we draw a direct analogy between the intricate, kpc-scale structures resolved here and those traced in face-on external galaxies. The solar neighborhood molecular gas exhibits a complexity that rules out simplistic geometric models. Moreover, the orientation of structures such as Cepheus Spur and the Sgr--Sct complex deviates significantly from the expected Galactic rotation, challenging models invoking purely rotational dynamics. Similarly, extragalactic analogs such as NGC 628 (M74) suggest that this matter accumulation likely results from a combination of Galactic shear and feedback-driven bubble expansion \citep{2023ApJ...944L..22B}, highlighting the need for high-resolution 3D mapping to uncover such features. 

\section{Conclusion}
\label{conclusion}

Based on the 2300 deg$^2$ $^{13}$CO data from the first phase of the MWISP survey along the northern Galactic plane, we constructed a novel $^{13}$CO cloud catalog. Utilizing Gaussian decomposition and velocity-component clustering, we identified over 30,000 cloud structures with at least 9 pixels, which are statistically optically thin to trace translucent molecular gas. For 3600 clouds with large angular sizes ($>43.5$ arcmin$^2$), we combined the BEEP methodology with the multi-jump point detection algorithm CU-JADE, incorporating high-precision \textit{Gaia} DR3 parallaxes to derive MC distances. The workflow effectively identifies cloud structures in velocity-crowded regions of the disk and detects multiple extinction jumps, and we successfully measured accurate distances for 1751 $^{13}$CO clouds. 

Based on this catalog (Table~\ref{table:tab1}), we draw the following conclusions:
\begin{enumerate}
    \item We present the largest $^{13}$CO cloud catalog with extinction-based distance measurements along the northern Galactic plane to date. The measured distances range from 139 pc to over 4 kpc. The fraction of the cataloged $^{13}$CO flux contained in clouds with accepted distances, is 42.26\% (37.31\% for the inner Galaxy). 
    
    \item Based on the CU-JADE's error propagation, distance uncertainties range from $\sim 2\%$ within 300 pc to $\lesssim 20\%$ beyond 2 kpc, establishing a clear reliability gradient. Accordingly, structures within 1 kpc constitute a `gold sample', while those within 2 kpc retain high reliability. 
    
    \item We obtain 25 $^{13}$CO cloud-maser pairs in position and radial velocity. The agreement between the extinction-based distances of the independently matched $^{13}$CO clouds and the maser parallaxes suggests a general trend in the distance accuracy for the retained sample. 
    
    \item We present the velocity-distance information of $^{13}$CO cloud structures in a ``scroll" format, serving as a compendium for molecular gas and star-forming regions. We provide the first extinction-based distance measurements for clouds associated with prominent regions such as DR21, W75N, and W44, etc. 
        
    \item We reveal several kpc-scale molecular structures in the solar neighborhood, which correspond to known structures, including the two kpc-scale superbubbles (the Oval Cavity at $l \sim 130^{\circ}$ and the Mon-Gem superbubble), the Local Spur (length $> 2$ kpc), and the Radcliffe wave, etc.
    
    \item In the face-on view, we identified several large-scale, four-dimensional gas structures. These include four sub-kpc scale giant cavities (the Cygnus Cavity A, Cygnus Cavity B, Vul-Aql Cavity, and Aql-Ser Cavity). Aql-Ser Cavity, the nearest, exhibits clear kinematic and 3D structural evidence, suggesting it is likely to be a channel connected to the Local Bubble.
    
    \item We resolved the substructure of the ``Split", and clarified the possible connection between Cepheus Spur, Cygnus Rift, and Cygnus-West. This reveals the ubiquity of spur-like structures within spiral arms and inter-arm regions.

    \item This work highlights the profound impact of stellar and supernova feedback, which influences the structural evolution and density distribution of the Galactic disk, potentially providing the necessary conditions to trigger future episodes of star formation.  
    
\end{enumerate}

The comprehensive catalog contains 1,751 cloud structures with reliably measured distances, enabling a quantitative characterization of the 3D distribution and kinematic properties of molecular gas in the solar neighborhood. The dataset thus provides a valuable observational basis for addressing key astrophysical questions, including the large-scale organization of the Galactic molecular disk, the kinematics and turbulence of MCs, the phase structure and chemical evolution of the ISM, and the connection between gas morphology and star formation activity. 

Initial decomposition of cloud structures and their 3D distribution within 5 kpc reveals a complex morphology in the solar vicinity, characterized by prominent large-scale features such as spurs, cavities, and coherent gas overdensities. These characteristics closely resemble those observed in face-on external spiral galaxies \citep[e.g.,][]{2023ApJ...944L..22B, 2023ApJ...944L..18M},  suggesting that the Milky Way’s molecular interstellar medium exhibits a comparable level of structural diversity. 
However, the characterization of cloud structures beyond approximately 1.5--2 kpc is increasingly limited by uncertainties in the Gaia parallaxes. This distance-dependent limitation is inherent to the current data and cannot be fully overcome by algorithmic improvements alone, although future Gaia data releases are expected to mitigate this limitation. 

Looking ahead, future Gaia DR4 data, combined with next-generation near-infrared astrometric missions, will substantially improve the precision and completeness of distance measurements, thereby extending the reliability of our catalog to larger distances and enabling the inclusion of more extended $^{12}$CO structures. Future work will also focus on deriving physical parameters (i.e., mass, velocity dispersion, and excitation temperature, etc.) for the full cloud sample, and on cross-correlating the catalog with HI, dust continuum, and infrared data to study the transitions between atomic and molecular gas phases. Comparison with state-of-the-art numerical simulations of the Galactic ISM will also provide critical tests of theoretical cloud formation and evolution models. Our work thus demonstrates the potential of combining large-scale CO surveys with precise astrometry, and provides a useful observational reference for future studies of the Milky Way’s molecular ISM as a dynamically and structurally complex system.

\begin{acknowledgments}

We appreciate the referee's constructive suggestions, which have significantly improved the quality of the paper. 
This research made use of the data from the Milky Way Imaging Scroll Painting (MWISP) project, which is a multiline survey in $\rm ^{12}CO$/$\rm ^{13}CO$/$\rm C^{18}O$ along the northern Galactic plane with the PMO 13.7 m telescope. We are grateful to all the members of the MWISP working group, particularly the staff members at the PMO-13.7m telescope, for their long-term support. MWISP was sponsored by the National Key R\&D Program of China with grants 2023YFA1608000, 2017YFA0402701, and the CAS Key Research Program of Frontier Sciences with grant QYZDJ-SSW-SLH047. Y.S. was supported by the National Natural Science Foundation of China (12673035, 12173090) and the Basic Research Program of Jiangsu (BK20252109). X.C. acknowledges the support  from the Tianchi Talent Program of Xinjiang Uygur Autonomous Region and the support by the CAS International Cooperation Program (grant No. 114332KYSB20190009). S.Z. acknowledges funding from the Specialized Research Fund for State Key Laboratory of Radio Astronomy and Technology (SDQZ-KF-A2612). This work also made use of data from the European Space Agency (ESA) mission Gaia (https://www.cosmos.esa.int/gaia), processed by the Gaia Data Processing and Analysis Consortium (DPAC, https://www.cosmos.esa.int/web/gaia/dpac/consortium). Funding for the DPAC has been provided by national institutions, in particular the institutions participating in the Gaia Multilateral Agreement. 
\end{acknowledgments}





%
\facilities{PMO: 13.7m}

\software{Astropy \citep{2013A&A...558A..33A,2018AJ....156..123A,2022ApJ...935..167A},  
          Scipy \citep{2020NatMe..17..261V}, 
          Matplotlib \citep{2007CSE.....9...90H}
          }


\appendix

\section{A sub-sample comparison with single-jump model} 
\label{appa}

\renewcommand\thefigure{\Alph{section}\arabic{figure}}
\setcounter{figure}{0}
\begin{figure*}[ht!]
\centering
\includegraphics[width=18cm,angle=0]{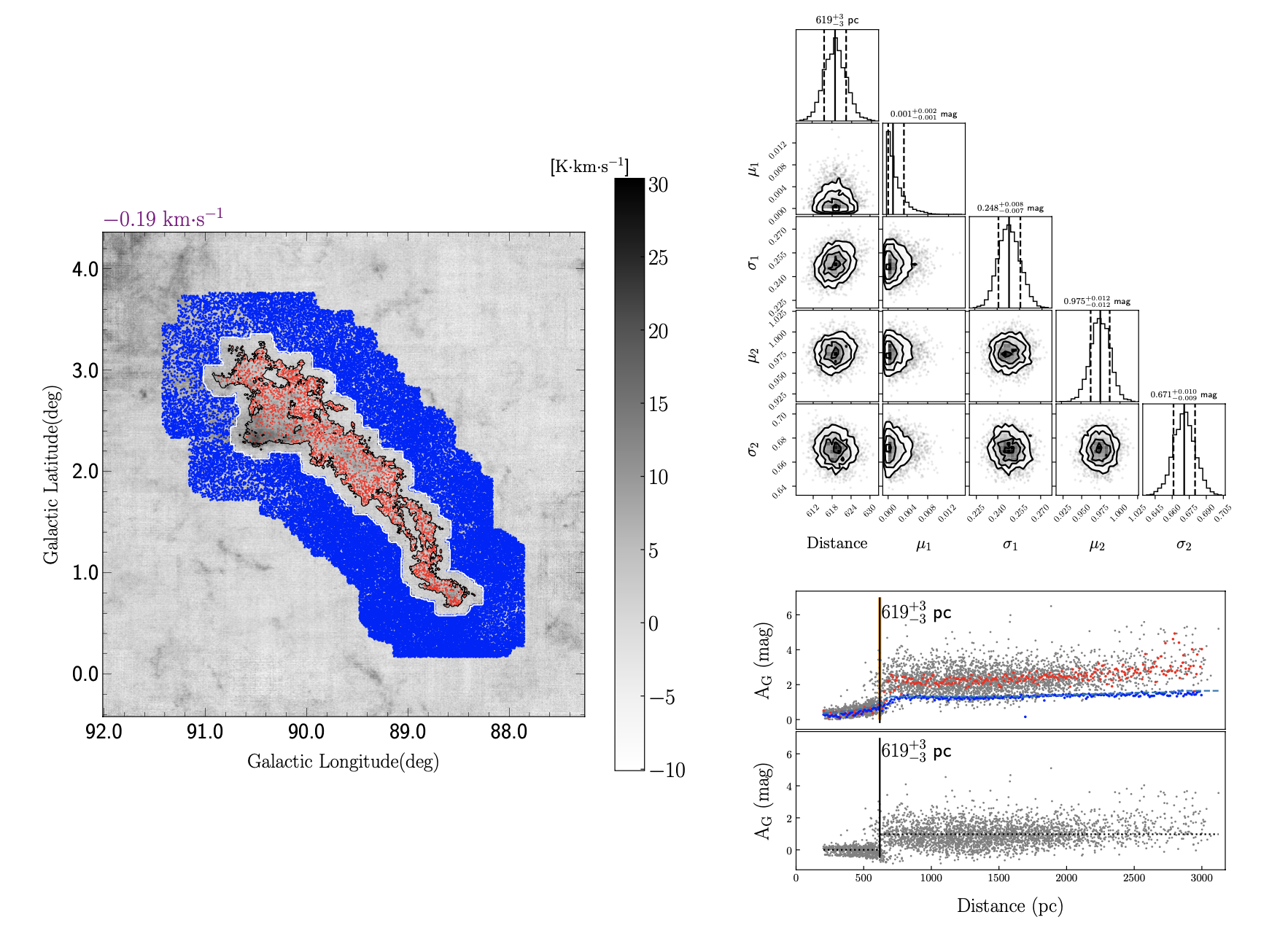}
\caption{An example for distance measurement using a simple one-jump-point model, same cloud as shown in Fig \ref{fig:fig2}. The upper right panel displays the posterior distributions for the fitted parameters, with the best-fitting values and their $1\sigma$ uncertainties indicated in the figure. \label{fig:figa1}}
\end{figure*}
\setcounter{figure}{0}

\renewcommand\thefigure{\Alph{section}\arabic{figure}}
\setcounter{figure}{1}
\begin{figure*}[h]
\centering
\includegraphics[width=10cm,angle=0]{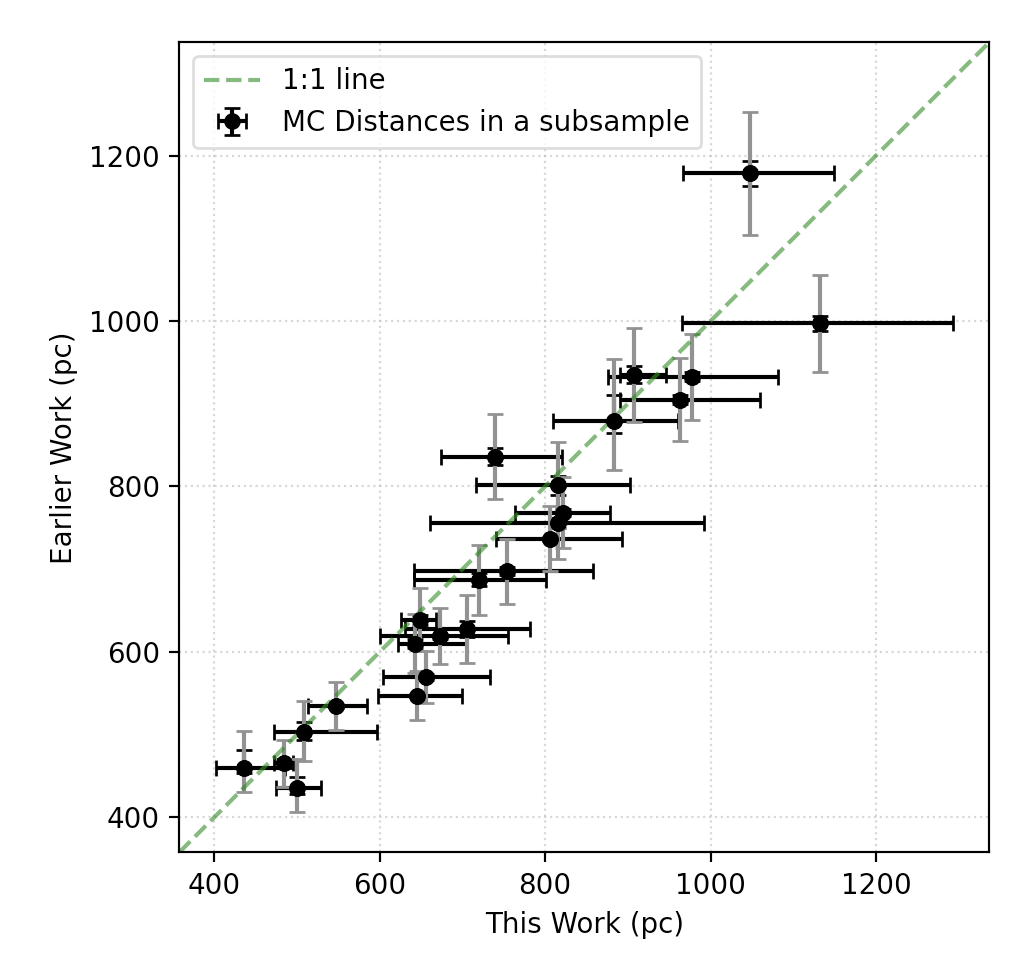}
\caption{Comparison of distance measurements between multi-jump and single-jump models within the same cloud identification framework.  \label{fig:figa2}}
\end{figure*}
\setcounter{figure}{1}

We compared the multi-jump point model CU-JADE with distance measurement methods from previous works (single-jump models) within the same cloud framework derived from Gaussian decomposition and guided by the BEEP principle. While both approaches perform similarly in the majority of steps, the primary difference lies in the determination of jump points. We selected a subsample of 26 large-scale clouds in Table \ref{table:tab1}. Among them, figure~\ref{fig:figa1} and Figure~\ref{fig:fig2} depict the same structure. To ensure stable background extinction fitting and MCMC sampling, a wider range of field stars was adopted here. Both methods reveal significant jump features, yet their resulting distances differ by $50 \sim 100$ pc. We attribute this discrepancy to the inherent ambiguity of the single-jump model. When multiple closely spaced jump points exist, the model tends to yield an averaged distance. Furthermore, the conflict between the simplistic single-jump assumption and the actual data complexity likely makes MCMC sampling under the Bayesian framework difficult to converge and computationally expensive.

As seen in Figure~\ref{fig:figa2}, aside from a few isolated clouds, a systematic offset of approximately 50 pc persists between the single-jump and multi-jump models. These closely spaced extinction components cannot be effectively removed by BEEP and contribute to increased distance uncertainties. Such deviations largely fall within the estimated distance uncertainties. Nevertheless, when comparing the two results, we added a 10\% systematic error \citep{2019ApJ...879..125Z} to the MCMC sampling uncertainties. Determining which method better reflects the true cloud distance requires more precise modeling of the 3D cloud structure. These closely spaced extinction components may represent distinct parts of a cloud or reflect its physical thickness. However, given their proximity, they cannot be distinguished with current parallax uncertainties. Higher-precision, larger-sample parallax measurements in the future are needed to resolve this issue.

\section{Comparison with DBSCAN-based Catalogs and Synergistic Presentation} \label{appb}

\renewcommand\thefigure{\Alph{section}\arabic{figure}}
\setcounter{figure}{0}
\begin{figure*}[ht!]
\centering
\includegraphics[width=12cm,angle=0]{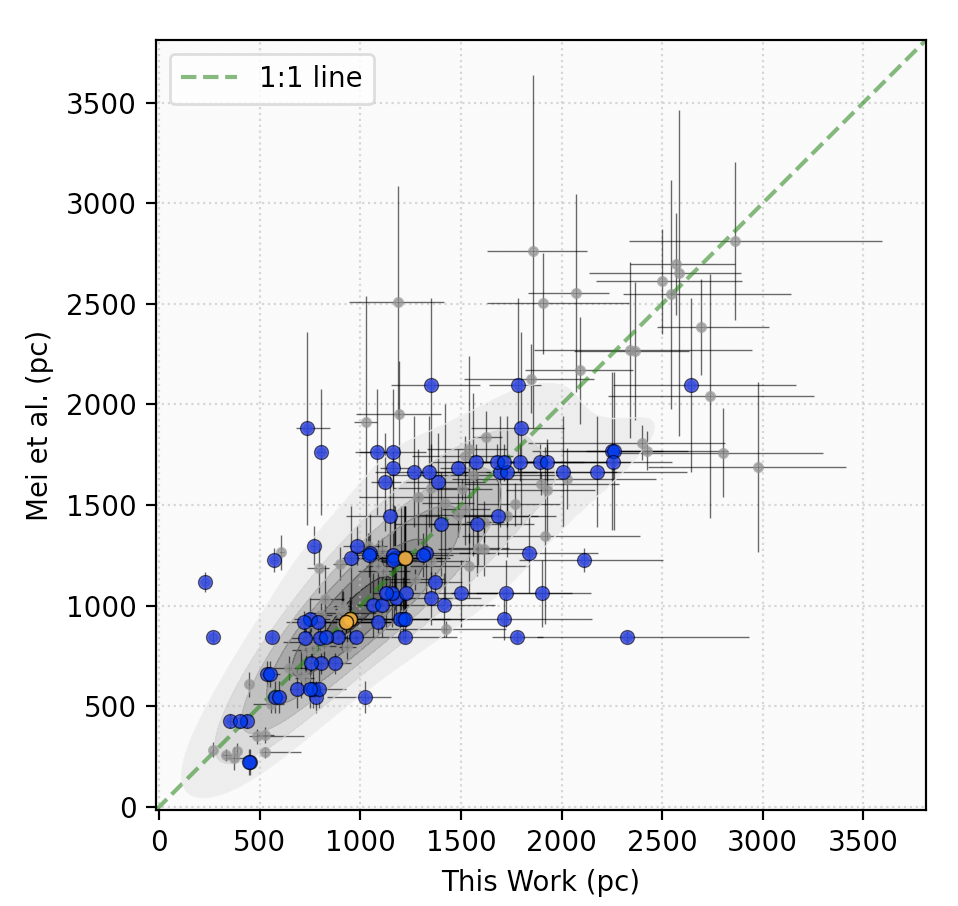}
\caption{Distance comparison after cross-matching with the \citet{2026ApJS..284...71M} cloud catalog, with 1$\sigma$ uncertainties shown for each data point. Grey points represent all 181 matched pairs. Blue points denote cases where a single DBSCAN-identified $^{13}$CO clump matches multiple $^{13}$CO structures identified in this work. Orange points indicate cases where a single $^{13}$CO structure from this work matches multiple DBSCAN-identified $^{13}$CO clumps.  \label{fig:figb1}}
\end{figure*}
\setcounter{figure}{0}

We compare our results with DBSCAN-based identifications to show that each method has its own strengths in different respects. Although DBSCAN performs well in multidimensional clustering, our Gaussian decomposition framework is better suited for isolating blended structures in velocity space. 

To compare the two algorithms, we prioritized using the same tracer, i.e., $^{13}$CO. Since distance measurements from \citet{2026ApJS..284...71M} are only available for a subset of the DBSCAN-identified $^{12}$CO clouds, we first matched the $^{13}$CO clumps identified by \citet{2026ApJS..282...65Y} using DBSCAN with the $^{13}$CO structures presented in this work. And we utilized the cross-index between $^{12}$CO and $^{13}$CO provided by \citet{2026ApJS..282...65Y} to conveniently trace back to the corresponding $^{12}$CO clouds and compare the distances. We employed a 3D Gaussian Mixture Model for the initial screening. Here, each DBSCAN clump was modeled as a 3D Gaussian distribution across ($l, b, v$) space. Subsequently, we performed a probabilistic matching of the clouds in Table~\ref{table:tab1} against these models, yielding 259 candidate pairs. Manual verification confirmed that 181 of these pairs correspond to the same physical structure.

It is important to note the presence of complex matching scenarios under different algorithms. In some cases, a single DBSCAN clump matched multiple clouds in Table~\ref{table:tab1} (blue points in Figure~\ref{fig:figb1}), implying that one $^{12}$CO cloud may encompass several distinct $^{13}$CO structures. Conversely, a single cloud in Table~\ref{table:tab1} was sometimes fragmented into different DBSCAN clumps, albeit belonging to the same $^{12}$CO cloud (orange points). 

Overall, the distances derived from the $^{12}$CO catalog agree well with those in Table~\ref{table:tab1}. However, significant outliers exist, particularly among the blue points. We attribute these discrepancies to three primary causes:
First, the DBSCAN catalog suffers from over-clustering, artificially grouping distinct structures at different distances. The horizontal sequence of blue points illustrates that a single $^{12}$CO cloud may contain multiple $^{13}$CO structures at various distances due to severe velocity crowding in the Galactic plane.
Second, fragmented $^{13}$CO structures exhibit larger distance measurement uncertainties, resulting in a broader distance spread. However, this effect alone cannot explain the extreme distance spans ($\sim 1$ kpc) observed for some clouds.
Third, $^{12}$CO emission may trace more diffuse, extended structures on ``super-cloud" scales. These structures possess significant depth along the LOS, whereas the method provides a single average distance, failing to capture the internal depth variation.

\renewcommand\thefigure{\Alph{section}\arabic{figure}}
\setcounter{figure}{1}
\begin{figure*}[ht!]
\centering
\includegraphics[width=18cm,angle=0]{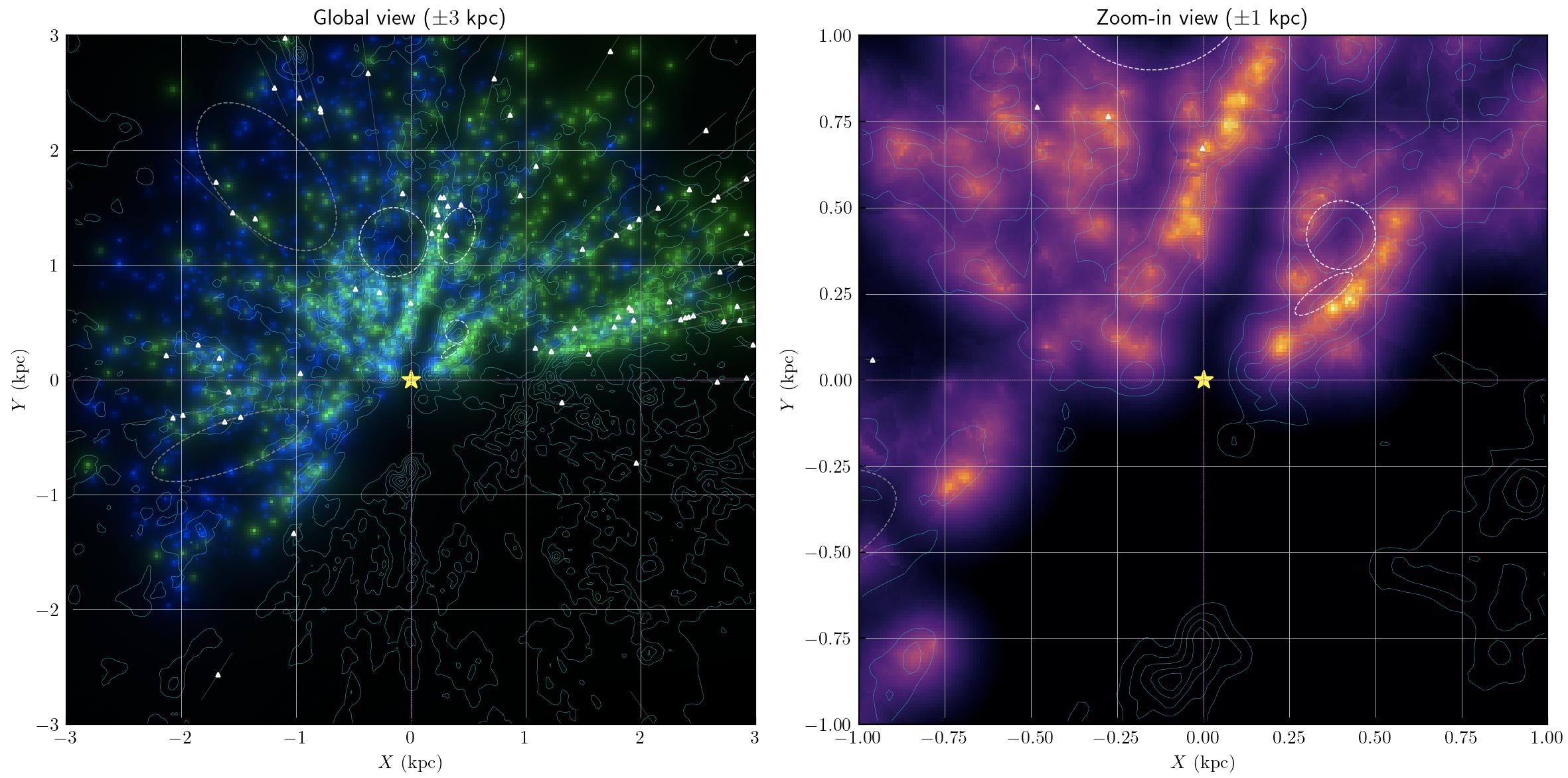}
\caption{\textit{Left panel:} Comparison between the equivalent surface density of the $^{13}$CO clouds listed in Table~\ref{table:tab1}, $^{12}$CO clouds in \citet{2026ApJS..284...71M}, and the 3D dust extinction map \citep{2022A&A...664A.174V} in the face-on view. Similar to Figure~\ref{fig:fig9} but a 3 kpc global view in left panel. The color scale represents the equivalent surface density derived from the gas tracer (green for $^{13}$CO and blue for $^{12}$CO), while the contours indicate the surface density derived from the dust tracer. The two tracers show excellent agreement in high-density regions but diverge in low-density areas. \textit{Right panel:} Zoomin view of equivalent surface density derived from $^{13}$CO cloud structures. The left panel shows the integration within $|z| < 400$ pc, and the right panel shows the integration within $|z| < 80$ pc.  \label{fig:figb2}}
\end{figure*}
\setcounter{figure}{1}

The different emphases of the two catalogs naturally lead to discrepancies in the derived cloud samples with distance estimates. We illustrate this using an equivalent density projection in Figure~\ref{fig:figb2}. The DBSCAN-based catalog (blue) places its primary emphasis on the outer Galaxy ($l > 90^{\circ}$), whereas our catalog reveals the 3D distribution of substantial molecular gas reservoirs within the inner Galaxy (green).

This disparity explains the limited number of matches discussed above, many large $^{12}$CO clouds could not be decomposed by DBSCAN, leading to failed distance measurements, while smaller $^{12}$CO clouds lacked significant $^{13}$CO emission. Consequently, the distance comparison is confined to an intermediate regime. This highlights the complementarity of the two catalogs: $^{12}$CO recovers more low-density, small-scale molecular gas, while the Gaussian decomposition-based $^{13}$CO catalog excels at resolving giant MCs in crowded regions. Together, they provide a comprehensive picture of the molecular gas distribution in the solar neighborhood.

\section{Foreground extinction, cloud clumpiness, and rejected distance measurements} \label{appc} 

Foreground extinction is an important selection effect for cloud catalogs based on Gaia distances, a point we have emphasized in our previous works \citep{2024AJ....167..220Z, 2025AJ....170..185Z}. Distant clouds toward the Galactic plane inevitably overlap with foreground structures along the LOS. When foreground MC extinction is strong (e.g., toward Cygnus and the Aquila--Serpens direction in the inner Galaxy), the foreground clouds manifest as Rift structures in the optical band. Since background Gaia stars are preferentially lost in the highest-extinction regions of a cloud, these directions exhibit an ``extinction wall'', leaving an insufficient number of background stars to support distance measurements to more distant clouds. 

At the same time, foreground MCs may exhibit hierarchical, porous structures. Both clumps and the low-extinction gaps between them can bias the distances to background MCs, leading to systematic underestimation. In Section~\ref{scroll}, we noted that the foreground MCs ($\lesssim 1$~kpc) in the Aquila--Serpens direction are extensive and elongated, and their distribution across low Galactic latitudes may introduce distance biases. To eliminate these biased measurements, which primarily arise from data limitations and sampling effects, we combined Larson's relations, the molecular gas scale height, the distance–velocity relation, and manual inspection in post-processing. 

Figure~\ref{fig:figc1} shows the distribution of Class~III samples, which have a concentration in the Aquila--Serpens region of the inner Galaxy. Figure~\ref{fig:figc1}(a) further reveals that most of the class III samples have large radial velocities ($|v_{\mathrm{LSR}}| \gtrsim 25~\mathrm{km~s^{-1}}$) and low Galactic latitudes, with widely extended, high-extinction foreground MCs along the LOS. We primarily used Figures~\ref{fig:figc1}(b) and (c) for inspection and rejection. The circled points represent Class~III samples that significantly deviate from Larson's size--linewidth relation. Owing to distance underestimation, these clouds appear smaller in size and their scale heights are close to 0~pc given their large velocity dispersions. In the distance-velocity relation, they appear as anomalous clusters that are nearby in distance but have relatively large radial velocities. We attribute these rejected distance measurements to the influence of foreground MCs and their internal structure: internal clumping and non-uniform extinction cause significant foreground extinction jumps even after off-cloud star selection and BEEP fitting for distant clouds. Since these clouds have large radial velocities and are likely distant, we consider their impact on the statistics of molecular gas flux in the solar neighborhood and on large-scale structures to be limited. 

Figure~\ref{fig:figc2} presents the $l$--$v$ distribution of MC structures toward the Aql--Ser cavity, where Class~III samples are concentrated. The left panel shows samples with precise distance measurements from Table~\ref{table:tab1}, and the right panel shows cloud structures from the full catalog within the same region. The overall inverted-V trend and velocity structure induced by the Aql--Ser cavity remain unchanged. First, all rejected samples (class III in $20^{\circ}<l<55^{\circ}$) have larger velocities ($\gtrsim 25~\mathrm{km~s^{-1}}$), thus they do not affect the low-velocity nearby structures. Second, Figure~\ref{fig:figc2} shows that the prominent low-velocity $\ell$--$v$ pattern toward Aquila--Serpens persists when clouds without distance estimates are included, supporting the projected velocity structure from the subsample with distance estimates. Some 96\% of clouds in the above $\ell$--$v$ range that lack distance solutions are smaller in angular size and not included in the parent sample. We consider these distance-unresolved small clouds (including some of the 1631 non-solution samples) to be algorithmically extracted fragments of larger MCs. Nevertheless, the locations of these distance-unresolved clouds along the LOS remain unknown, so their effect on the three-dimensional cavity boundaries is uncertain. Given that the distance-unresolved clouds contribute only a minor fraction (flux ratio $\lesssim16\%$) of the distance-measured sample and trace the same projected velocity structure, their inclusion may not significantly blur the large-scale cavity boundaries. Finally, we suggest that the sample with missing or rejected distance measurements has at most a minor effect on the interpretation of nearby, low-velocity large-scale structures.

\renewcommand\thefigure{\Alph{section}\arabic{figure}}
\setcounter{figure}{0}
\begin{figure*}[ht!]
\centering
\includegraphics[width=13.8cm,angle=0]{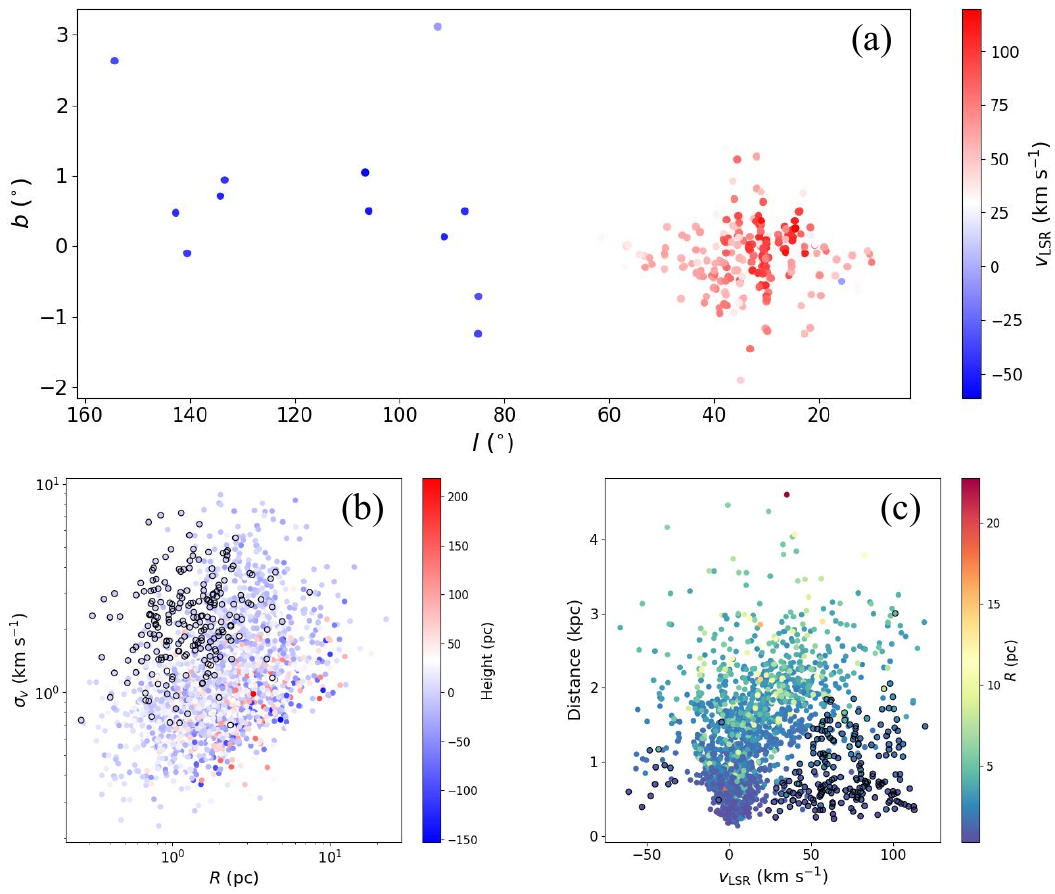}
\caption{Distribution of the 218 Class~III samples in $l$--$b$ and parameter spaces.
(a) Scatter points show the projected positions of the Class~III cloud structures, with colors indicating radial velocity.
(b) Distribution of cloud structures in the velocity dispersion--size plane. Class~III samples are circled, and colors indicate the height above the Galactic plane at the measured distance.
(c) Distribution of cloud structures in the velocity--distance plane. Class~III samples are circled, and colors indicate the physical size of the cloud at the measured distance.  \label{fig:figc1}}
\end{figure*}
\setcounter{figure}{0}

\renewcommand\thefigure{\Alph{section}\arabic{figure}}
\setcounter{figure}{1}
\begin{figure*}[ht!]
\centering
\includegraphics[width=18cm,angle=0]{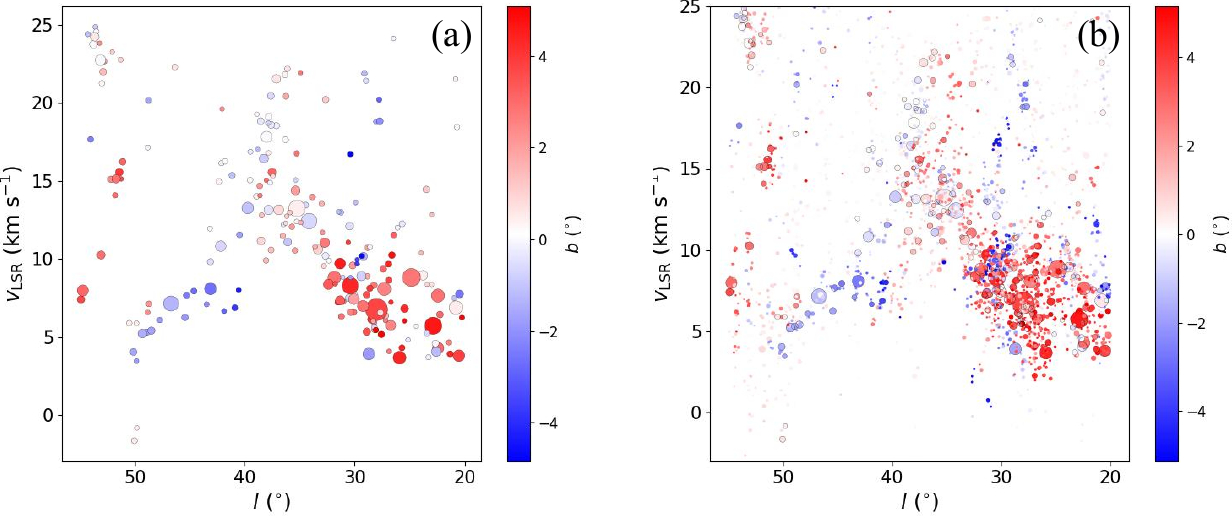}
\caption{Longitude--velocity ($l$--$v$) distribution of $^{13}$CO cloud structures
toward the Aquila--Serpens direction.
Left panel (a): cloud structures with precise distance measurements from Table~\ref{table:tab1},
with colors indicating Galactic latitude and point sizes proportional to angular size.
Right panel (b): distribution of cloud structures from the full catalog within the same region,
with colors also indicating Galactic latitude and point sizes consistent with those in the left panel.
Clouds with distance measurements are circled.  \label{fig:figc2}}
\end{figure*}
\setcounter{figure}{1}


\bibliography{ngpi}{}
\bibliographystyle{aasjournalv7}



\end{document}